\documentclass[11pt]{article}

\usepackage[utf8]{inputenc}
\usepackage[T1]{fontenc}
\usepackage{lmodern}
\usepackage[a4paper,margin=1in]{geometry}
\usepackage{microtype}
\usepackage[round]{natbib}
\usepackage{listings}
\usepackage{doi}

\usepackage{amsmath, amssymb, amsthm, stmaryrd}

\usepackage{algorithm}
\usepackage{algpseudocode}

\usepackage[dvipsnames]{xcolor}
\definecolor{DarkRed}{rgb}{0.6,0,0}
\usepackage{hyperref}
\hypersetup{
  colorlinks=true,
  linkcolor=blue,   
  citecolor=blue,   
  urlcolor=DarkRed  
}
\usepackage{orcidlink}

\usepackage{graphicx}
\usepackage{booktabs}

\usepackage{framed}      
\usepackage{fancyvrb}    

\newcommand{\proglang}[1]{\textsf{#1}}  
\newcommand{\pkg}[1]{\textbf{#1}}       
\newcommand{\code}{\lstinline}          

\newcommand{\Keywords}[1]{\par\noindent\textbf{Keywords: }#1.}

\newcommand{\A}{\mathcal{A}}
\DeclareMathOperator{\lk}{\llbracket}
\DeclareMathOperator{\rk}{\rrbracket}
\DeclareMathOperator{\I}{\mathbb{I}}
\newcommand{\Ss}[2]{\displaystyle\sum_{#1}^{#2}}
\def\bZ{\mathbb{Z}}
\def\cA{\mathcal{A}}
\def\bR{\mathbb{R}}
\def\cS{\mathcal{S}}
\def\bP{\mathbb{P}}
\def\bE{\mathbb{E}}
\DeclareMathOperator{\Prob}{\mathsf{P}}

\DefineVerbatimEnvironment{CodeInput}{Verbatim}{fontsize=\small}
\DefineVerbatimEnvironment{CodeOutput}{Verbatim}{fontsize=\small}

\newenvironment{CodeChunk}{\vspace{0.5em}}{\vspace{0.5em}}

\newcommand{\AuthorBlock}[2]{%
  \begin{minipage}[t]{.48\linewidth}\centering
    \Large \textbf{#1}\\[0.5em] 
    \large #2                   
  \end{minipage}%
}

\title{\pkg{hdMTD}: An \proglang{R} Package for High-Dimensional Mixture Transition Distribution Models}

\author{%
  \AuthorBlock{%
    Maiara Gripp\,\orcidlink{0009-0003-5852-6531}%
  }{%
    Universidade Federal\\ do Rio de Janeiro%
  }%
  \hfill
  \AuthorBlock{%
    Giulio Iacobelli\,\orcidlink{0000-0002-9591-8388}%
  }{%
    Universidade Federal\\ do Rio de Janeiro%
  }%
  \\[5em]
  \AuthorBlock{%
    Guilherme Ost\,\orcidlink{0000-0003-0887-9390}%
  }{%
    Universidade Federal do\\ Rio de Janeiro%
  }%
  \hfill
  \AuthorBlock{%
    Daniel Y. Takahashi\,\orcidlink{0000-0003-4972-001X}%
  }{%
    Universidade Federal\\ do Rio Grande do Norte%
  }%
}

\date{\today}

\begin{document}
\maketitle

\begin{abstract}
Several natural phenomena exhibit long-range conditional dependencies. High-order mixture transition distribution (MTD)  are parsimonious non-parametric models to study these phenomena. An MTD is a Markov chain in which the transition probabilities are expressed as a convex combination of lower-order conditional distributions. Despite their generality, inference for MTD models has traditionally been limited by the need to estimate high-dimensional joint distributions. In particular, for a sample of size $n$, the feasible order $d$ of the MTD is typically restricted to $d \approx \mathcal{O}(\log{n})$. To overcome this limitation, Ost and Takahashi (2023) recently introduced a computationally efficient non-parametric inference method that identifies the relevant lags in high-order MTD models, even when $d \approx \mathcal{O}(n)$, provided that the set of relevant lags is sparse. 
In this article, we introduce \pkg{hdMTD}, an \proglang{R} package allowing us to estimate parameters of such high-dimensional Markovian models.
Given a sample from an MTD chain, \pkg{hdMTD} can retrieve the relevant past set using the BIC algorithm or the forward stepwise and cut algorithm described in Ost and Takahashi (2023). The package also computes the maximum likelihood estimate for transition probabilities and estimates high-order MTD parameters through the expectation-maximization algorithm. Additionally, \pkg{hdMTD} also allows for simulating an MTD chain from its stationary invariant distribution using the perfect (exact) sampling algorithm, enabling Monte Carlo simulation of the model. 
We illustrate the package's capabilities through simulated data and a real-world application involving temperature records from Brazil.
\end{abstract}

\Keywords{mixture transition distributions, Markov chains, high-dimension, \proglang{R}, perfect sample}

\section{Introduction}
Categorical time series with long-range dependencies are ubiquitous in nature. Today's weather depends not only on the weather of the previous day, but also on what the weather was like a year ago. 
In large language models, incorporating relevant words from distant positions in the input sequence can substantially enhance next-word prediction (more than $10^5$ for ChatGPT 4o). Economic indicators' dynamics depends on events that happen at different temporal scales, including long-range dependencies. Modeling these phenomena using stochastic processes with possibly long-range dependencies is natural. Given the complex nature of some of these phenomena, often non-parametric modeling is desirable. Nevertheless, non-parametric modeling, like high-order Markov chains, generally requires estimating joint probability distributions, which imposes significant constraints on the size of the past that can be included in the model before the estimation becomes inaccurate. To ameliorate the problem, \citet{Raftery85} introduced the mixture transition distribution (MTD) model, a subclass of finite-order Markov chains, as a parsimonious non-parametric model for categorical time series. The MTD framework successfully reduces the number of parameters required for estimation, alleviating the complexity associated with high-order dependencies. Despite this advancement, accurately estimating time series dependencies of an MTD model that stretch far into the past remained limited.
Recently, \citet{Ost&Takahashi}  introduced a computationally and statistically efficient solution to the problem of estimating time-series long-range dependencies if the time series is a realization of an MTD model in which the number of relevant lags in the past is not too big.  
This opened the possibility to make non-parametric inferences for the class of high-order MTD models. 
In this article, we introduce \pkg{hdMTD}, an \proglang{R} package allowing us to estimate parameters of such high-dimensional Markovian models.

The core of the \pkg{hdMTD} package lies in the algorithms that, given a sample from an MTD chain, can retrieve the relevant  set of past lags using two different methods: (1) the Bayesian information criterion (BIC) algorithm, which is a classical model selection method; (2) the forward stepwise and cut (FSC) algorithm, which is a computationally and statistically efficient algorithm for estimating the set of relevant lags even when the temporal distance between the lags is close to the sample size. Once the relevant lags are specified, \pkg{hdMTD} package can also compute the maximum likelihood estimate (MLE) of the transition probabilities and estimate high-order MTD parameters using the expectation-maximization (EM) algorithm. 
To allow for Monte Carlo experiments, it is necessary to sample from the invariant distribution of the process. For low-order Markov chains, traditionally we rely on running the process with fixed pasts and wait until the process approximate the invariant measure. When the order of the chain is large, this approach can be inefficient as it is difficult to decide whether the joint probability measure is close the to the invariant distribution. A more robust approach is to sample exactly from the invariant distribution. Algorithms that allow to draw samples from the invariant distribution of a process are called perfect sampling algorithms (see \citep{Fernandez2001Coupling} for a comprehensive introduction on the subject). Our package allows to simulate MTD chains of any order from its invariant distribution using the perfect sampling algorithm, which is novel to the best of our knowledge .

A brief review of related computational packages helps to place \pkg{hdMTD} in context. Markovian models have a long history and broad applications, nevertheless, practical tools for inference, particularly for mixture transition distribution (MTD) models, remain relatively limited. Traditional statistical software (e.g., \proglang{SPSS} \citep{SPSS}, \proglang{Stata} \citep{Stata}) do not provide native, dedicated support for Markov models, while \proglang{SAS} \citep{SAS} offers limited functionality, primarily for homogeneous chains. Consequently, researchers increasingly rely on open-source environments like \proglang{R} \citep{R-base} for advanced Markovian analysis.
In \proglang{R}, the \pkg{march} \citep{march} \citep{Berchtold2020} package provides dedicated tools for MTD model estimation, offering likelihood based optimization for categorical stochastic processes. This package has become a reference for MTD applications, supporting homogeneous Markov chains, hidden Markov models, and other related architectures.  Its companion package \pkg{GenMarkov} \citep{GenMarkov} extends this to multivariate MTDs. Notably, none of these tools provide perfect simulation algorithms of MTD processes given parameters. 
For general Markov chain analysis, the \pkg{markovchain} \citep{markovchain17} package analyzes properties (e.g., stationarity) and experimentally fits $k$-step Markov chains, while \pkg{DTMCPack} \citep{DTMCPack} focuses on simulation from user-defined matrices. The \pkg{depmixS4} \citep{depmixS4} package offers tools for fitting standard and hidden Markov models with both continuous and categorical responses. Outside \proglang{R}, specialized tools like \proglang{Stata}’s \code{markov} module \citep{Statamarkov} implement basic discrete-time Markov chain analysis. More recently, developments in \proglang{Julia} \citep{Julia} and \proglang{Python} \citep{Python} have started to offer basic Markovian model support, but tools specific for MTDs remain largely unavailable.

We end the introduction with a description of the structure of the paper. In Section \ref{sec:modeldef} we introduce the main notation and define the MTD model. In Section \ref{sec:methods} we formulate the statistical lag selection problem and describe two solutions for it. We briefly describe the perfect sampling algorithm in Section \ref{sec:perfect_simulation}. Finally, in Section \ref{sec:usingMTD} we describe the \pkg{hdMTD} package with its functions and usages. We also exhibit examples of applications using a simulated data and an analysis of daily temperature data in Brazil spanning several years.

\section{Model definition} \label{sec:modeldef}
\subsection{General notation}
We denote $\mathbb{Z}=\{\dots,-1,0,1,\dots\}$ the set of integers, $\mathbb{Z}^+=\{1,2,\dots\}$ the set of positive integers and $\mathbb{R}$ the set of real numbers.
For $t,s\in\mathbb{Z}$ with $t> s$, we write $\lk s,t \rk$ to denote the discrete interval $\{s,s+1,\dots, t-1,t\}$.
Throughout the text, $\cA$ denotes a finite subset of $\bR$.
For $S\subset\bZ$, we denote $\mathcal{A}^S$ the set of all $\cA$-valued sequences $x_S=(x_j)_{j\in S}$ indexed by the set $S$. When $S=\lk -d, -1\rk$ for some $d\in\bZ_+$, we omit the subscript $S$ from the sequence $x_S$, denoting such sequence simply by $x$, and write $\cA^d$ instead of $\cA^{\lk -d, -1\rk}$ to alleviate the notation.
For two probability measures $\mu$ and $\nu$ on $\cA$,
\begin{align*}
    d_{TV}(\mu,\nu)=\frac{1}{2}\Ss{a\in\A}{}|\mu(a)-\nu(a)|\;,
\end{align*}
denotes the {\it total variance distance} between $\mu$ and $\nu$.

\subsection{Markov models}
We say that a discrete-time stochastic chain $\{X_t\}_{t\in\mathbb{Z}}$ taking values in the {\it state space} $\cA$ is a \textit{Markov chain of order} $d\in\mathbb{Z}^+$, if for all state $a\in\A$, time  $t\in\mathbb{Z}$ and $k\in\bZ^{+} $ past states $x_{-k},\ldots, x_{-1}\in \cA$ satisfying $\mathbb{P}(X_{t-k}=x_{-k},\ldots, X_{t-1}=x_{-1})>0$ with $k>d$, it holds that  
\begin{equation}
\label{def:markov_property}
  \mathbb{P}(X_t=a|X_{t-1}=x_{-1},\ldots, X_{t-k}=x_{-k})=\mathbb{P}(X_t=a|X_{t-1}=x_{-1},\ldots X_{t-d}=x_{-d})\,.
\end{equation}
We say that a Markov chain is \textit{stationary}, if for all $t,s,h \in\mathbb{Z}$ with $t\geq s$ and $x_s,\ldots, x_t\in\cA$,
$$\mathbb{P}(X_{s}=x_s,\ldots, X_{t}=x_t)=\mathbb{P}(X_{s+h}=x_s,\ldots, X_{t+h}=x_{t})\,.$$
In what follows, we write $\pi$ to denote the distribution of a stationary Markov chain. For any vector $x\in\cA^d$ of $d$ past states, we write $\pi(x)$ to denote $\mathbb{P}(X_{-d}=x_{-d},\ldots, X_{-1}=x_{-1})$ and $\pi(x_S)$ to denote $\mathbb{P}(X_{j}=x_{j},j\in S)$ for $S\subset \lk -d,-1\rk $ and $x_S\in\A^S$. 

For a stationary Markov chain, the conditional probabilities in \eqref{def:markov_property} do not depend on the time index $t\in\bZ$. In that case, we denote for $S\subseteq \lk -d, -1\rk$, $t\in\bZ$, $a\in\cA$ and $x_S\in\cA^S$, 
$$
\Prob(a|x_S)=\bP\left(X_t=a|X_{t+j}=x_j,j\in S\right)\,.
$$
The set $\{\Prob(\cdot|x):x\in\cA^d\}$ is called the family of {\it transition probabilities} of the Markov chain.
Throughout this article, we consider only stationary Markov chains.

\subsection{Mixture transition distribution (MTD) models}\label{MTDsec}

An MTD \citep{Raftery85,Berchtold&Raftery}  of order $d\in\mathbb{Z}^+$ is a Markov chain of order $d$ with state space $\A$, where the transition probabilities can be expressed as
\begin{align}
    \Prob(a|x)=\lambda_0p_0(a)+\Ss{j=-d}{-1}\lambda_jp_j(a|x_j)\,,\ a\in\A\,, \ x\in\cA^d\;,\label{eq:MTD}
\end{align}
with $\lambda_0,\lambda_{-1},\dots,\lambda_{-d}\in [0,1]$ satisfying $\Ss{j=-d}{0}\lambda_j=1$. Moreover $p_0(\cdot)$ and $p_j(\cdot|b)$, $j\in\lk -d,-1\rk$ and $b\in\A$, are probability measures on $\A$. 

Henceforth, an index $j\in\lk -d,0\rk$ will be called the \textit{j-th lag} of the model. Throughout this article, for $j\in \lk-d,-1\rk$, we denote by $p_j=\{ p_j(a|b):\ a,b\in\A\} $, the $|\A|\times|\A|$ \textit{stochastic matrix} representing the relationship between the $j$-\textit{th} lag and the present. Equation \eqref{eq:MTD} represents what \citet{Berchtold&Raftery} called a \textit{multimatrix MTD model}, since the stochastic matrices $p_j$ are allowed to be different from each other. 
Note that a multimatrix MTD model may have up to $d+(|\A|-1)(1+d|\A|)$ parameters. Hence, for such Markovian models the number of parameters grows linearly with the order $d$, unlike general Markov chains for which the number of parameters grows exponentially with the order. 
Note also that there are even more parsimonious MTD models, such as the \textit{single matrix MTD}, where all lags influence the distribution of the present state equally (i.e., given any $a,b\in\A$, $p_j(a|b)=p(a|b)$ for all $j\in\lk -d,-1\rk$). This latter MTD model has only $d+(|\A|-1)(1+|\A|)$ parameters.

Equation \eqref{eq:MTD} has the following interpretation.
To sample a state for the chain at any given time, we can first sample a lag $j\in\lk -d,0\rk$ with probability $\lambda_j$. If $j=0$, we can proceed by sampling a state in $\A$ according to the distribution $p_0$, which does not depend on the past states. However, if a lag $j\in \lk -d,-1\rk$ is chosen (which can only occur if $\lambda_j>0$), then we must sample a state from the conditional distribution $p_j(\cdot|x_j)$. This distribution depends on the $d$ past states $x\in \cA^d$, but only through the state at lag $j$ (i.e., the value $x_j$). 

For any Markov chain of order $d$, we may measure the relevance of a lag $j$ through the \textit{oscillation} $\delta_j$, defined as
\begin{align}
    \delta_j=\max\big\{d_{TV}(\Prob(\cdot|x),\Prob(\cdot|y)):x,y \in\A^{d} \ \text{such that} \ x_{k}=y_{k}, \ \forall \ k\in \lk -d,-1\rk\setminus\{j\}\big\}\,.
    \label{eq:osc}
\end{align}
Plugging \eqref{eq:MTD} in \eqref{eq:osc}, one can check that for an MTD model of order $d$, the oscillation $\delta_j$ can be rewritten as
\begin{align}
    \delta_j=\lambda_j\cdot\underset{b,c\in\A}{\max}\big\{d_{TV}\big(p_j(\cdot|b),p_j(\cdot|c)\big)\big\}\,.
    \label{eq:oscMTD}
\end{align}
From \eqref{eq:oscMTD}, we see that if either $\lambda_j=0$ or $\underset{b,c\in\A}{\max}\{d_{TV}(p_j(\cdot|b),p_j(\cdot|c))\}=0$, then the oscillation $\delta_j=0$. In that case, the lag $j$ is irrelevant to the transition probabilities of the chain in the sense that $\Prob(a|x)$ is a constant function of $x_j$ for all $d$ past states $x\in\cA^d$.
In light of this, we define
\begin{align}
    \Lambda=\{j\in\lk-d,-1\rk: \delta_j>0\}\,,
\end{align}
the \textit{set of relevant lags}. 
From the definition of the set $\Lambda$, one can show that $\Prob(a|x)=\Prob(a|x_S)$ for all $x\in\cA^d$, $a\in\cA$ and $S\subseteq \lk-d,-1\rk$ containing $\Lambda$.
Therefore, finding this set is essential for obtaining the sparsest representation of the dependence on past states in an MTD model.  

 \section{Problem formulation and methods} \label{sec:methods}
\subsection{Statistical lag selection}

Suppose we are given a sample $\cS_n:=(X_1,\ldots, X_n)$ from an MTD model of known order $d < n$ and unknown set of relevant lags $\Lambda$. In the statistical lag selection problem, the goal is to estimate the set of relevant lags $\Lambda$ from the sample $\cS_n$. In what follows, we briefly present the four different estimators for the set of relevant lags that were implemented in the \pkg{hdMTD} package. The first one is the Bayesian information criterion (BIC) estimator described in Section \ref{BICsec}. The CUT estimator, presented 
in Section \ref{CUTsec}, is the second.  
The forward stepwise (FS) and forward stepwise and cut (FSC) estimators are the other two, both being described in Section \ref{FSsec}. 
Before presenting these four estimators, we need first to introduce the empirical conditional probabilities. This is done in the Section \ref{subsec:est_condi_proba} below.

\subsection{Empirical probabilities}
\label{subsec:est_condi_proba}

In the \pkg{hdMTD} package, given a sample $\cS_n$ from an MTD model of known order $d\in\mathbb{Z}^+$, we 
compute $\hat{\pi}_{n,j}(x_S, b, a)$, an estimator of the joint probability of observing a symbol $a\in\A$ at lag $0$, a sequence $x_S\in \A^S$ at lags in $S\subseteq\lk-d,-1\rk$, and a symbol $b\in\A$ at lag $j\in\lk-d,0\rk\setminus S$, as
\begin{align}\label{eq:piest}
    \hat{\pi}_{n,j}(x_S, b, a) = \dfrac{N_{n,j}(x_S,b,a)}{n-d}\;,
\end{align} where
\begin{align*}
N_{n,j}(x_S,b,a) = 
\sum_{t=d+1}^{n} \mathbb{I} \left\{X_{t+k} = x_k, k\in S, X_{t+j}=b, X_t = a \right\}\;,
\end{align*} is the number of times the sequence $x_S$, at lags in $S$, appears in the sample along with symbols $b$ at lag $j$ and $a$ at lag $0$. Note that for $j=0$, unless $b=a$, we have $N_{n,j}(x_S,b,a)=0$. When $a=b$, we write $N_{n}(x_S,a) = N_{n, 0}(x_S,a,a)$ and $\hat{\pi}_{n}(x_S,a) = \hat{\pi}_{n, 0}(x_S,a,a)$ for brevity. 

Using these estimates, we compute the estimator of the conditional probability of observing a symbol $a \in \A$ after a sequence  $x_S \in \A^S$ at lags in $S \subseteq \lk -d, -1 \rk$, and symbol $b\in \A$ at lag $j\in \lk-d,0\rk\setminus S$, as
\begin{align*}
    \hat{\Prob}_{n,j}(a|x_S, b)=
    \begin{cases} 
\dfrac{\hat{\pi}_{n,j}(x_S,b,a)}{\hat{\pi}_{n,j}(x_S,b)}, &  \text{if} \ \hat{\pi}_{n,j}(x_S,b)>0 \\
\\
|\A|^{-1}, & \text{otherwise }
\end{cases},
\end{align*}\footnote{Note that if $j=0$, then $\hat{\Prob}_{n,j}(a|x_S,b)$ corresponds to the empirical probability of $X_t=a$ given that $X_t=b$. Therefore, if $\hat{\pi}_{n}(x_S,b)>0$, then $\hat{\Prob}_{n,0}(a|x_S,b)=\I\{b=a\}$. }
where $\hat{\pi}_{n,j}(x_S,b) = \Ss{a\in\A}{} \hat{\pi}_{n,j}(x_S, b, a)\;.$ 
We also compute
\begin{align*}
    \bar{N}_n(x_S) = \sum_{a \in \A} N_n(x_S,a) \text{ and }     \hat{\pi}_n(x_S) = \sum_{a\in\A} \hat{\pi}_n(x_S,a)\;,
\end{align*}
which, respectively, count and estimate the frequency with which the sequence  $x_S$ appears in the first $n-1$ symbols of the sample (i.e., in $(X_1,\ldots, X_{n-1})$) and
\begin{align}
    \label{eq:qxa}
    \hat{\Prob}_{n}(a|x_S)=
    \begin{cases} 
\dfrac{\hat{\pi}_{n}(x_S,a)}{\hat{\pi}_{n}(x_S)}, &  \text{if} \ \hat{\pi}_{n}(x_S)>0\\
\\
|\A|^{-1}, & \text{otherwise}
\end{cases}\;,
\end{align}
which is an estimator of the conditional probability of observing a symbol $a \in \A$ at lag $0$  after the sequence $x_S \in \A^S$ at lags in $S$.

One can show that, when $\Lambda\subseteq S$, the empirical conditional probability $\hat{\Prob}_n(a|x_S)$ defined in \eqref{eq:qxa} converges in probability to the conditional probability $\Prob(a|x_{\Lambda})$ as the sample size $n\to\infty$,
provided that  
any two vectors $x,y\in\cA^d$ of $d$ past states communicate, meaning that the MTD model generating the sample is irreducible (see, e.g.,  \citet[Chapter 3]{Finesso}).
Under this irreducibility assumption and if
$\hat{\Lambda}_n=\hat{\Lambda}_n(\cS_n)$ is a consistent estimator of $\Lambda$ (i.e., the probability of the event $\{\hat{\Lambda}_n\neq \Lambda\}$ vanishes as the sample size $n\to\infty$), then we also have
\begin{align*}
    \hat{\Prob}_n(a|x_{\hat{\Lambda}_n})\rightarrow \Prob(a|x_{\Lambda})\,,\ \text{in probability as }n\rightarrow\infty\,,
\end{align*}
for all $a\in\mathcal{A}$ and $x_{\Lambda}\in\cA^{\Lambda}$ such that $\pi(x_{\Lambda})>0$. Similar results can also be proved for the empirical conditional probabilities $\hat{\Prob}_{n,j}(a|x_{S},b)$.

\subsection{BIC estimator}\label{BICsec}

Given a sample $\cS_n$, a positive integer $d\in\mathbb{Z}^+$ and a subset $S^*\subseteq \lk-d,-1\rk$, the BIC estimator returns a subset, denoted $\hat{\Lambda}_n^{\text{BIC}}(\cS_n)$, achieving the minimum of a penalized log-likelihood criterion over all non-empty subsets $S \subseteq S^*$. 
More precisely, the BIC estimator is defined as 
\begin{align}
\label{eq:BIC}
    \hat{\Lambda}_n^{\text{BIC}}(\cS_n)=\underset{S\subseteq S^*}{\arg\min}\Ss{a\in\A}{}\Ss{\substack{x_{S}\in\A^S}}{}-N_n(x_S,a)\log \hat{\Prob}_n(a|x_S)+ \text{Pen}(S,n)\,,
\end{align}
where $\text{Pen}(S,n)=\theta(S)\log(n)\xi$ is the penalty term, with $\xi>0$ being a constant to be chosen and $\theta(S)$ denoting the number of parameters in the model. 
In an MTD model with $\lambda_0>0$, $\theta(S)=|S|+(|\A|-1)(1+|\A|\zeta)$  where $\zeta\in\{1,\dots,|S|\}$ is the number of \textit{distinct matrices} $p_j$.
In the \pkg{hdMTD} package, the function that computes the BIC estimator is called \code{hdMTD_BIC}.

Model selection through the BIC \citep{Schwarz} is a well-known method. The difference between BIC values of two models approximates twice the logarithm of the Bayes factor between them. The Bayes factor represents the evidence, provided by the data, in favor of one model against another \citep{Kass&Raftery} and is known for its validity for comparison of multiple and non-nested models. Also, the BIC has been proven to be a consistent method for estimating the order of Markov chains \citep{Katz,Csiszar&Shields}. 

One drawback of the BIC estimator is its high computational complexity when $|S^*|$ is large. For example, suppose that $S^*=\lk-d,-1\rk$. 
In this case, according to \eqref{eq:BIC}, any subset $S\subseteq \lk-d,-1\rk$ of size $1\leq l\leq d$ is a candidate set for the true set of relevant lags and, therefore, the number of candidate sets is $\sum_{l=1}^{d}\binom{d}{l}=2^d-1$ (exponential in $d$). Hence, computing the BIC estimator becomes impractical for large values of $d$ (e.g. $d=n\beta$ for some $\beta\in (0,1)$) without any prior knowledge on the set of relevant lags. 
When some prior information about the set of relevant lags is available, for instance if the user knows $|\Lambda|=l$ in advance, the number of candidates sets is then reduced to $\binom{d}{l}$ (polynomial in $d$).

\subsection{CUT estimator}\label{CUTsec}

\citet{Ost&Takahashi} proposed a consistent estimator for the set of relevant lags of an MTD model based on pairwise comparisons of empirical conditional probabilities corresponding to \textit{compatible} pasts. In this article, we call it CUT estimator\footnote{In \citet{Ost&Takahashi}, this estimator based on pairwise comparisons is called PCP estimator.}.

Given a subset $S\subseteq \lk -d,-1\rk$ and a lag $j\in S$, two sequences $x_S,y_S\in\A^S$ are called $(S\setminus \{j\})$-\textit{compatible} if $x_{k}=y_{k}$ for all $k\in S\setminus\{j\}$.
From our discussion at the end of Section \ref{MTDsec}, if $j\in\Lambda\subseteq S$ there must exist a pair of $(S\setminus \{j\})$-compatible pasts $x_S,y_S\in \A^S$ such that the total variation distance between $\Prob(\cdot|x_S)$ and $\Prob(\cdot|y_S)$ is strictly positive. Hence, the CUT estimator runs across all $j\in S$, and tests if there is a pair of $(S\setminus \{j\})$-compatible pasts $x_S,y_S\in \A^S$ for which 
\begin{align}
d_{TV}\big(\hat{\Prob}_n(\cdot|x_S),\hat{\Prob}_n(\cdot|y_S)\big)>t_n(x_S,y_S)\,,
    \label{eq:critCut}
\end{align}
where $t_n(x_S,y_S)$ denotes a threshold value depending on $n, x_S$ and $y_S$ (as well as on some other tuning parameters) that is explicitly defined below.
Given a lag $j \in S$, if no pair of $(S \setminus \{j\})$-compatible pasts satisfies \eqref{eq:critCut}, then lag $j$ is not included in the estimated set of relevant lags $\hat{\Lambda}_n^{\rm CUT}=\hat{\Lambda}_n^{\rm CUT}(\cS_n)$.

For pasts $x_S,y_S\in \A^S$, the threshold that appears in \eqref{eq:critCut} is defined as $t_n(x_S,y_S)=s_n(x_S)+s_n(y_S)$, where $s_n(x_S)$ is given by
\begin{align*}
s_n(x_S)=\sqrt{\dfrac{\alpha(1+\xi)}{2\bar{N}_n(x_S)}}\Ss{a\in\A}{}\sqrt{\dfrac{\mu}{\mu-\psi(\mu)}\Big(\hat{\Prob}_n(a|x_S)+\dfrac{\alpha}{\bar{N}_n(x_S)}\Big)}+\dfrac{\alpha|\A|}{6\bar{N}_n(x_S)}\,,
\end{align*}
with $\xi>0$, $\alpha>0$ and $\mu\in(0,3)$ such that $\mu>\psi(\mu)=e^{\mu}-\mu-1$. The threshold $t_n(x_S,y_S)$
has been derived through martingale concentration inequalities (see \citet[Appendix B]{Ost&Takahashi}). An interesting feature of this threshold is that it is 
adapted to each realization of sample $\cS_n$ of the MTD model, i.e., it may change depending on the realization of the sample. Using uniform thresholds that do not depend on the sample typically lead to statistical procedures that
either underestimate or overestimate the set of relevant lags.   

The CUT estimator is computed in the function \code{hdMTD_CUT}. In \citet{Ost&Takahashi}, a few theoretical guarantees of the CUT estimator have been established. Let us highlight two of them. First, the CUT estimator can be implemented with $O(|\cA|^2|S|(n-d))$ operations (see \citet[Remark 2-(c)]{Ost&Takahashi}). Second, in \citet[Corollary 4]{Ost&Takahashi}, the CUT estimator is shown to be consistent even when the order $d$ is large (e.g. $d=n\beta$ for some $\beta\in (0,1)$),
provided that $\Lambda\subseteq S$ and
$|S|$ is relatively small with respect to the sample size $n$ (i.e., $|S|=O(\log(n))$). However, this assumption is too restrictive and might be difficult to be justified in practice. To circumvent this issue, one can use the FS or FSC estimators described subsequently.

\subsection{FS and FSC estimators}\label{FSsec}

The FS estimator \citep{Ost&Takahashi} is defined iteratively as follows. At each step, it receives a subset $S\subseteq \lk-d,-1\rk$ as input and computes, based on the sample $\cS_n$, the empirical estimate $\hat{\nu}_{n,j,S}$ of the quantity
$
\bar{\nu}_{j,S}=\bE(\nu_{j,S}(X_S)),
$
for each lag $j\in S^c$. Here, for each $x_S\in\cA^S$,
\begin{align*}
   \nu_{j,S}(x_S)=\frac{1}{2}\sum_{a\in\cA}\sum_{b\in\cA}\sum_{c\in\cA}w_{j,S}(b,c,x_S)|\bP_{x_S}(X_0=a|X_j=b)-\bP_{x_S}(X_0=a|X_j=c)|\,, 
\end{align*}
where $w_{j,S}(b,c,x_S)=\bP_{x_S}(X_j=b)\bP_{x_S}(X_j=c)$. The quantity $\nu_{j,S}(x_S)$ measures the influence of the lag $j$ on the distribution of the current state, given that the states at the lags in $S$ equal $x_S$. Averaging this conditional influence over the possible pasts $x_S$ gives the quantity $\bar{\nu}_{j,S}$, which is estimated by 
\begin{align}\label{eq:hatnu}
    \hat{\nu}_{n,j,S}=\Ss{x_S\in\A^S}{}\Ss{b\in\A}{}\Ss{c\in\A}{}\dfrac{\hat{\pi}_{n,j}(x_S,b)\hat{\pi}_{n,j}(x_S,c)d_{TV}\Big(\hat{\Prob}_{n,j}(\cdot|x_S,b),\hat{\Prob}_{n,j}(\cdot|x_S,c)\Big)}{\hat{\pi}_n(x_S)}.
\end{align}
Once the values of the empirical estimates $\hat{\nu}_{n,j,S}$ for all $j\in S^c$ are computed, a new lag $j^*$ is then included to the subset $S$, where 
\begin{align*}
    j^*=\underset{j\in S^c} {\arg\max}\ \hat{\nu}_{n,j,S}\,.
\end{align*}
At the next step, the above procedure is repeated starting now from the enlarged subset $S\cup\{j^*\}$. At the first step, we take $S=\emptyset$. The total number of steps $\ell \in \mathbb{Z}^+$ used to compute the FS estimator has to be chosen by the user.  
In particular, the set $\hat{\Lambda}_n^{\rm FS}$ built from the FS estimator has size $\ell$. In the \pkg{hdMTD} package, the function that computes the FS estimator is called \code{hdMTD_FS}.

As indicated in \citet[Remark 8 - (c)]{Ost&Takahashi}, one can compute the FS estimator with $O(|A|^3\ell(n -d)(d -(\ell - 1)/2)$ operations. Moreover,   
\textit{Theorem 2} of \citet{Ost&Takahashi} states that, on a certain event and for a suitable chosen $\ell$, we have that $\Lambda \subseteq \hat{\Lambda}_n^{\rm FS}$. In \pkg{hdMTD} package, $\ell$ is a user input, and in practice, we verify that if $\ell \geq |\Lambda|$, then $\Lambda \subseteq \hat{\Lambda}_n^{\rm FS}$ with high probability.  
This suggests that the FS output might be a reasonable choice for input set $S$ of both BIC and CUT estimators.
Let us mention that using the FS output as the CUT estimator input constitutes what \citet{Ost&Takahashi} refer to as the FSC estimator. 
Roughly speaking, the FSC estimator is proved to be consistent even when the underlying MTD model has a large order $d$, as long as the size of the set of relevant lags $\Lambda$ grows at most logarithmically with the sample size $n$. As opposed to the CUT estimator, 
the FSC estimator does not require any prior knowledge of a sufficiently small subset $S$ containing the set of relevant lags $\Lambda$, which is a huge advantage in practice.

In order to establish the consistency of the FSC estimator, it is essential to split the sample between the FS and CUT estimators, ensuring that the same portion of the sample is not used twice.
Hence, to compute the FSC estimator, we first split the data into two parts. The first part of the sample is used to compute the FS estimator. Setting the output of the FS estimator as the input set $S$, we then compute the CUT estimator using the second part of the sample.    
In the \pkg{hdMTD} package, the function for computing the FSC estimator is called \code{hdMTD_FSC}.

\section{Algorithm to exactly simulate MTD models}
\label{sec:perfect_simulation}

The \pkg{hdMTD} package also provides an {\it exact} or {\it perfect} simulation algorithm for an MTD model, i.e., an algorithm that outputs states distributed {\it exactly} as the stationary distribution of the MTD model. 
As mentioned in the introduction, this can be particularly useful in the context of Mont Carlo estimation when the underlying MTD model has high order. 

The algorithm can be described briefly as follows. Suppose we want to simulate $X_t$, the state of the MTD model at time $t$. To that end, we consider independent random variables $L_s$, with $s\leq t$, taking values in $\lk 0, d\rk$ in such a way that $\bP(L_s=j)=\lambda_{-j}$, for all $s\leq t$. In the first step of the algorithm, we look at the value of $L_t$. If $L_t=0$, then we set $X_t\sim p_0(\cdot)$ (i.e., the value of $X_t$ is chosen according to the distribution $p_0(\cdot)$)  and stop the algorithm. Otherwise, we have that $L_t>0$ and we go to the step 2 of the algorithm, where we look at the value of $L_s$ where $s=t-L_t$. If $L_s=0$, we set $X_s\sim p_0(\cdot)$ and $X_{t}\sim p_{-L_t}(\cdot|X_s)$, and then stop the algorithm. Otherwise, we go to step 2 and repeat. One can show that the algorithm stops after a finite number of steps almost surely whenever $\lambda_0>0$. Moreover, the value $X_t$ produced by the algorithm is distributed according to the stationary distribution of the MTD model given in \eqref{eq:MTD}.  
Furthermore, we can generate a sample $\cS_n$ from an MTD model under its stationary distribution by applying the above algorithm to each variable $X_t$, $1\leq t\leq n$, starting from $X_n$, reusing the values of the random variables $X_s$ that have already been generated if necessary. 
The pseudocode of our perfect simulation is given in Algorithm~\ref{pseudo:code}. 

\begin{algorithm}[hbt!]
\setcounter{algorithm}{0}
\caption{Exact simulation: \texttt{perfectSample(MTD, N)}}
\label{pseudo:code}
\textbf{Input:} an MTD object and the intended sample size $N$\\
\Comment{ \textit{an MTD object is parameterized by $\Lambda \subset \mathbb{Z}^-$, $\A$, $\lambda_0$, $p_0(\cdot)$, $\{\lambda_j, p_j(\cdot | x_j), x_j \in \A: j \in \Lambda\}$}}\\
\textbf{Output:}  a vector $X$ of size $N$, sampled from the MTD's invariant distribution

\Comment{ \textit{main function}}
\begin{algorithmic}[1]
\Function{perfectSampleStep}{$t$}
\State Sample $j$ from $\Lambda \cup \{0\}$ with probability $\lambda_j$
\If{$j = 0$}
    \State Sample $X_t$ from $p_0(\cdot)$
\Else
    \If{$X_{t+j}$ is not yet sampled}
        \State \Call{perfectSampleStep}{$t + j$}
    \EndIf
    \State Sample $X_t$ from $p_j(\cdot | X_{t+j})$
\EndIf
\EndFunction
\end{algorithmic}

\Comment{\textit{iteration}}

\begin{algorithmic}[1]
\For{$t = -N$ to $-1$}
    \State \Call{perfectSampleStep}{$t$}
\EndFor
\State Return $X_{-1}, \dots, X_{-N}$
\end{algorithmic}
\end{algorithm}

There is no exact simulation algorithm implemented in any of the packages for MTD models that we are aware of, namely, the packages \pkg{march} and \pkg{GenMarkov}. 
In the \pkg{hdMTD}, the above exact simulation algorithm is implemented in the function \code{perfectSample}. 
Our algorithm is inspired by \citep{Comets2002Proceses} - where the authors present an algorithm to exactly simulate processes with long memory. In that paper, the authors also discuss the differences and similarities of their algorithm with respect to other classical exact simulation algorithms such as {\it coupling from the past} algorithm introduced by \citep{Propp1996Exact}.    
We also refer to \citep{Fernandez2001Coupling} for a comprehensive introduction to perfect simulation algorithms.
\newpage

\section[Using hdMTD]{Using \pkg{hdMTD}}\label{sec:usingMTD}
In this section, we show how to use the \pkg{hdMTD} package through illustrative examples. 
The package can be installed in \proglang{R} with the following command:
\begin{CodeChunk}
\begin{CodeInput}
R> install.packages("hdMTD") 
\end{CodeInput}  
\end{CodeChunk}
and loaded with:
\begin{CodeChunk}
\begin{CodeInput}
R> library("hdMTD") 
\end{CodeInput}  
\end{CodeChunk}

\subsection{Data generation}

Suppose we want to obtain a sample from an MTD model with state space $\cA=\{0,1\}$ where the set of relevant lags is $\Lambda=\{-30,-15,-1\}$. First, we need to specify such a model. We do so with the \code{MTDmodel} function. This function creates a \textbf{class MTD object} whose attributes are all the necessary parameters for defining an MTD model, and the resulting transition matrix. 
The only inputs that are absolutely necessary are the \textbf{set of relevant lags} (\code{Lambda}) and the \textbf{state space} (\code{A}). Note that while $\Lambda\subset \mathbb{Z}^-$, the argument \code{Lambda} of the function \code{MTDmodel} must be in $\mathbb{Z}^+$. Hence, the input set \code{Lambda} has to be interperted as the set $-\Lambda$.
The user can also input the MTD \textbf{weights} (a number \code{lam0} representing $\lambda_0$ and a vector \code{lamj} representing the values $\lambda_j$ for all $j\in\Lambda$), the \textbf{past conditional distributions} (a \code{list} \code{pj} where each entry is a \proglang{matrix} representing the matrices $p_j$, $j\in\Lambda$), and/or the \textbf{independent distribution} (a vector \code{p0}, representing $p_0(a)$, $a\in \A$). If not provided, these parameters are sampled uniformly and normalized\footnote{For sampling $p_0(a)$, $a\in\A$, the function first sets \code{p0<-runif(length(A))}, and then renormalizes the vector $p_0$ to obtain a probability measure by defining \code{p0<-p0/sum(p0)}. The same procedure is done for the \textbf{weights} (\code{lam0} and \code{lamj}). For sampling $p_j(a|b)$, $a,b\in\A$, $j\in\Lambda$, the function sets a \code{list} called \code{pj} with $|\Lambda|$ matrices, where each row of each matrix is sampled in the same way as \code{p0}.
}. 
\begin{CodeChunk}
\begin{CodeInput}
R> set.seed(11)
R> Lambda <- c(1, 15, 30)
R> A <- c(0, 1)
R> lam0 <- 0.01
R> lamj <- c(0.39, 0.3, 0.3)
R> p0 <- c(0.5, 0.5)
R> MTD <- MTDmodel(Lambda = Lambda, A = A, lam0 = lam0, lamj = lamj, p0 = p0)
R> summary(MTD)
\end{CodeInput}
\begin{CodeOutput}
Mixture Transition Distribution (MTD) model 

Call:
MTDmodel(Lambda = Lambda, A = A, lam0 = lam0, lamj = lamj, p0 = p0)

Relevant lags: -1, -15, -30
State space: 0, 1

lambdas (weights):
  lam0  lam-1 lam-15 lam-30 
  0.01   0.39   0.30   0.30 

Independent distribution p0:
p0(0) p0(1) 
  0.5   0.5 

Transition matrices pj (one per lag):
 
 pj for lag j = -1:
           0         1
0 0.35190318 0.6480968
1 0.03558321 0.9644168
 
 pj for lag j = -15:
          0         1
0 0.4278830 0.5721170
1 0.7670555 0.2329445
 
 pj for lag j = -30:
          0         1
0 0.8341439 0.1658561
1 0.2184814 0.7815186

Transition matrix P: 8 x 2
- Preview of first rows of P:
            0         1
000 0.5208503 0.4791497
001 0.3974855 0.6025145
010 0.6226020 0.3773980
011 0.4992372 0.5007628
100 0.3361516 0.6638484
101 0.2127868 0.7872132

Reading guide for P:
Rows list past contexts from oldest to newest, matching lags (-30, -15, -1).
\end{CodeOutput}
\end{CodeChunk}
The first element of the vector \code{lamj} must refer to the smallest element in \code{Lambda} (the largest element in $\Lambda$), the second one corresponds to the second smallest element of \code{Lambda} and so forth. In the example \code{lamj = c}($\lambda_{-1}$, $\lambda_{-15}$, $\lambda_{-30}$).  Regarding the vector \code{p0}, its first element should correspond to the smallest element in \code{A} and so on. In our case,  \code{p0=c(}$p_0(0)$\code{,}$p_0(1)$\code{)}. If the MTD model does not have the independent distribution $p_0$, set the argument \code{indep_part} to \code{FALSE} in the \code{MTDmodel} function. In this case, \code{p0} will not be sampled and \code{lam0} is automatically set to $0$. 
Observe that we omitted the argument \code{pj} in the \code{MDTmodel} function, yet a list with 3 stochastic matrices is generated (sampled uniformly and then normalized). Alternatively, we could have specified \code{pj = list('p-1' = matrix(\dots), 'p-15' = \ matrix(\dots), 'p-30' = matrix(\dots))}, defining the matrices associated to the relevant lags. To replicate the same matrix for all lags set the argument \code{single_matrix} to \code{TRUE}. After specifying the parameters of the model, the matrix \code{P} is computed through the convex sum in \eqref{eq:MTD}.

In addition to the \code{summary()} method, the \code{MTD} object includes a set of accessor functions and methods that facilitate model exploration and manipulation. Accessor functions such as \code{transitP()}, \code{lambdas()}, \code{pj()}, \code{p0()}, \code{lags()}, \code{Lambda()}, and \code{states()} return the core components of the model in a structured way. The object also supports various S3 methods, including \code{coef()}, \code{logLik()}, \code{probs()}, \code{perfectSample()}, and \code{oscillation()}---the last two being further discussed later in this section.
To guide the user, the \code{print()} method displays a concise overview of the model structure, along with the full list of available accessors and methods.

Once we have an MTD model, we can sample from its \textbf{invariant distribution} using the \\\code{perfectSample} function, provided that \code{lam0 > 0}.

\begin{CodeChunk}
\begin{CodeInput}
 R> X <- perfectSample(MTD, N = 1000) 
\end{CodeInput}
\end{CodeChunk}

where \code{N} is the intended sample size, and \code{MTD} is a \proglang{class} MTD object. If \code{lam0 = 0} an \proglang{error} will occur since \code{perfectSample}'s algorithm requires \code{lam0 > 0} to run, as discussed in Section \ref{sec:perfect_simulation}.

\subsection{Estimation}

With the sample at hand, we proceed to the problem of estimating the relevant lags using the \code{hdMTD} functions. The \pkg{hdMTD} package includes four distinct estimators for the \textbf{set of relevant lags}: the BIC, FS, CUT and FSC estimators.  An overview of these estimators is provided in Section \ref{sec:methods}. Users can either use the specific functions \code{hdMTD_"method"} (e.g., \code{hdMTD_FSC()}) or the more general \code{hdMTD()} function, indicating the method via the \code{method} argument (e.g., \code{method = "FSC"}). 
The specific functions provide a more direct interface, with tailored documentation and argument prompts that guide the user more effectively. The \code{hdMTD()} function, on the other hand, returns a structured object with additional information, which can be useful for further inspection.
However, note that the arguments required by \code{hdMTD()} depend on the chosen method and are not directly visible.

\begin{CodeChunk}
    \begin{CodeInput}
R> hdMTD_FS(X, d = 40, l = 4)        
    \end{CodeInput}
    \begin{CodeOutput}
[1] 30 15  1 27
    \end{CodeOutput}
\end{CodeChunk}

In the example above, we use the function \code{hdMTD_FS} to obtain, within the first 40 pasts (hence \code{d = 40}), those \code{l = 4} pasts that are most relevant according to the FS estimator\footnote{The FS estimator initially identifies the past that has the most significant predictive power for the present. Then, if $l>1$, it iteratively searches for the lag that is most important given the knowledge of the previous lags, and so on. In our output, the lag -1 was found to be the most relevant, followed by lag -30 in conjunction with the information from lag -1, and so forth.}.

Alternatively, the same estimation can be performed using the general \code{hdMTD()} function with \code{method = "FS"}, which returns an object with additional information:

\begin{CodeChunk}
    \begin{CodeInput}
R> FS <- hdMTD(X, d = 40, method = "FS", l = 4)
R> S(FS); summary(FS)
    \end{CodeInput}
    \begin{CodeOutput}
[1] 30 15  1 27
hdMTD lag selection

Call:
hdMTD(X = X, d = 40, method = "FS", l = 4)

Method: FS
Order upper bound (d): 40
Selected S set: 30, 15, 1, 27

Relevant lag set estimated by FS method : -30, -15, -1, -27
    \end{CodeOutput}
\end{CodeChunk}

The \code{hdMTD()} function returns an object of class \proglang{hdMTD}, which stores the estimated lags along with auxiliary information. The selected lags can be accessed using the accessor \code{S()} (output in $\mathbb{Z}^+$) or \code{lags()} (output in $\mathbb{Z}^-$), and additional details are available via the \code{summary()}. To include the full list of arguments used during estimation, use \code{summary(x, settings = TRUE)}. In the example above, the estimated set contains the true lags $-30$, $-15$, and $-1$. Since \code{l = 4}, the algorithm also included one additional lag ($-27$ in this case). Had we used \code{l = 3}, the function would have returned exactly \code{Lambda}, given that $27$ was the last output.

Now, let us see how the other estimators work.
In what follows, we illustrate each method using its specific function (e.g., \code{hdMTD_BIC}), which provides a direct output of the selected lags. Alternatively, all estimators can also be accessed through the general \code{hdMTD()} interface by specifying the corresponding method.
We can look for the 4 relevant pasts of the MTD model that minimizes \eqref{eq:BIC} with the \code{hdMTD_BIC} function.

\begin{CodeChunk}
    \begin{CodeInput}
R> hdMTD_BIC(X, d = 40, minl = 4, maxl = 4)
    \end{CodeInput}
    \begin{CodeOutput}
[1]  1 15 17 30
    \end{CodeOutput}
\end{CodeChunk}

In this example, the BIC method computes the penalized log-likelihood of all possible models with 4 relevant pasts between 1 and 40 (there are $\binom{40}{4}$ such models). Among all combinations of 4 lags, the combination of lags $-30,-17,-15,-1$ has the smallest penalized log-likelihood.
Note that from this output alone, we cannot determine which of the 4 returned lags is more important (differently from the output of \code{hdMTD_FS}).

Another difference is that the running of BIC estimator took about 30 minutes,
while the FS method took less than 7 seconds.
If we have more information about the set of relevant lags, for example, suppose that we know that the relevant lags belong to the subset $S=\{1,5,10,15,17,20,27,30, 35,40\}\subset \lk 1:d\rk$, we may pass this information to the function through the argument \code{S}.
\begin{CodeChunk}
    \begin{CodeInput}
R> hdMTD_BIC(X, d = 40, S = c(1, 5, 10, 15,  17, 20, 27, 30, 35, 40),
+       minl = 4, maxl = 4)
    \end{CodeInput}
    \begin{CodeOutput}
[1]  1 15 17 30
    \end{CodeOutput}
\end{CodeChunk}
Now \code{hdMTD_BIC} searched only across all subsets of \code{S} of size 4 (since \code{minl=maxl=4}). Given that $\{1, 15, 17, 30\}\subset S$ has size 4, the output is the same as in the unconstrained case. However, the computation time is much faster (less than 10 seconds). We can also allow the function to look for sets with other   sizes by changing \code{minl} and \code{maxl} arguments.
\begin{CodeChunk}
    \begin{CodeInput}
R> hdMTD_BIC(X, d = 40, S = c(1, 5, 10, 15, 17, 20, 27, 30, 35, 40),
+       minl = 1, maxl = 4)
    \end{CodeInput}
    \begin{CodeOutput}
[1]  30
    \end{CodeOutput}
\end{CodeChunk}
The \code{hdMTD_BIC} function iterates over all possible subsets with size 1 (in this case, there were 10 of them), then over all possible subsets of size 2, and so on, until reaching all possible subsets of size 4 (since \code{minl = 1, maxl = 4}). In this scenario, the model with single relevant past $-30$ is the most likely given the sample and the penalization criterion (recall that BIC penalizes the number of parameters). Since the model with pasts $-30$, $-17$, $-15$, and $-1$ was within the possibilities, this shows that, given the sample, the BIC criterion selects a more parsimonious model, for the increase in likelihood did not outweigh the loss from the extra parameters. The penalization term is multiplied by a constant which can be passed to the model with the argument \code{xi}, by default, \code{xi = 0.5}. Changing this constant may alter the output, as a larger constant increases the penalty term, while a smaller constant decreases it. The \code{hdMTD_BIC} function can return, for each size (i.e., from size \code{minl} to \code{maxl}), the most likely set, by setting the argument \code{byl = TRUE}. Setting \code{BICvalue} to \code{TRUE}, makes the function  return the BIC value of each selected set.
\begin{CodeChunk}
    \begin{CodeInput}
R> hdMTD_BIC(X, d = 40, S = c(1, 5, 10, 15, 17, 20, 27, 30, 35, 40),
+       minl = 1, maxl = 4, byl = TRUE, BICvalue = TRUE)
    \end{CodeInput}
    \begin{CodeOutput}
          30        15,30      1,15,30   1,15,17,30 smallest: 30 
    644.4959     648.0111     649.4950     650.2869     644.4959
    \end{CodeOutput}
\end{CodeChunk}

With this output we can see the BIC values of the most likely MTD models, each having a set of lags of size ranging from \code{minl} to \code{maxl}. The smallest is the single past $-30$ model (with a BIC value of 644.4959). Now let us decrease just a little our penalization constant, from $0.5$ to $0.4$:
\begin{CodeChunk}
    \begin{CodeInput}
R> hdMTD_BIC(X, d = 40, S = c(1, 5, 10, 15, 17, 20, 27, 30, 35, 40),
+       minl = 1, maxl = 4, byl = TRUE, BICvalue = TRUE, xi = 0.4)
    \end{CodeInput}
    \begin{CodeOutput}
      30    15,30   1,15,30  1,15,17,30 smallest: 1,15,17,30 
641.7328 643.1757  642.5873  641.3069               641.3069 
  
    \end{CodeOutput}
\end{CodeChunk}
With this change in the constant, the penalty across more parameters was reduced and the model with four lags returned the smallest BIC value.

Note that, by default, our \code{hdMTD_BIC} function assumes the sample is generated from an MTD model where, for all $j,k\in\Lambda$, there exists at least one symbol $b\in\A$ for which $p_j(\cdot|b)\neq p_k(\cdot|b)$. This represents a \textit{multimatrix} MTD where all the $p_j$ matrices are different, hence a model having the largest number of parameters possible (see Section \ref{BICsec}). However, if the user sets the argument \code{single_matrix} to \code{TRUE}, the function will consider, for any $a,b\in\A$ and all $j\in\Lambda$, $p_j(a|b)=p(a|b)$, a \textit{single matrix} MTD with fewer parameters and hence, a smaller penalization term. The user can also set \code{indep_part} to \code{FALSE} and the \code{hdMTD_BIC} function will assume $\lambda_0=0$, which also lessens the penalization term.

\begin{CodeChunk}
    \begin{CodeInput}
R> hdMTD_BIC(X, d = 40, S = c(1, 5, 10, 15, 17, 20, 27, 30, 35, 40),
+       minl = 1, maxl = 4, byl = TRUE, BICvalue = TRUE,
+       single_matrix = TRUE, indep_part = FALSE)
    \end{CodeInput}
    \begin{CodeOutput}
      30    15,30   1,15,30  1,15,17,30 smallest: 1,15,17,30 
637.5881 634.1956  628.7718  622.6559               622.6559
    \end{CodeOutput}
\end{CodeChunk}

Since we did not specify a value for \code{xi}, the function \code{hdMTD_BIC} uses the default value of \code{xi = 0.5}. This time we have set \code{indep_part} to \code{FALSE} and \code{single_matrix} to \code{TRUE}. Hence, the \code{hdMTD_BIC} function considers the model to be a \textit{single matrix} MTD with $\lambda_0=0$ when calculating the penalization term. In a \textit{single matrix} MTD model, having a large number of relevant lags has much less impact on the penalization term when compared to a \textit{multimatrix} MTD. As a result, in this example, the model with smallest BIC value has more relevant lags than when compared to the second to last example.

Given a sample, the CUT estimator verifies if the relevance of each lag is larger than a certain threshold, which depends on some parameters \code{alpha} ($\alpha$), \code{mu} ($\mu$) and \code{xi} ($\xi$) that can be passed to the \code{hdMTD_CUT} function (see Section \ref{CUTsec}). If these parameters are not specified, the function \code{hdMTD_CUT} uses \code{alpha = 0.05, mu = 1, xi = 0.5} by default. The user must input the order \code{d} ($d$) of the model, and a vector \code{S}, representing the input subset $S\subseteq \lk1,d\rk$. If \code{S} is not provided as an input, the function uses \code{S = 1:d}. After testing each lag in \code{S} for its relevance, the function \code{hdMTD_CUT} returns the ones that were not removed or `cut'.
\begin{CodeChunk}
    \begin{CodeInput}
R> hdMTD_CUT(X, d = 40, S = c(1, 5, 10, 15, 17, 20, 27, 30, 35, 40))
    \end{CodeInput}
    \begin{CodeOutput}
[1] 1  5 10 15 17 20 27 30 35 40
    \end{CodeOutput}
\end{CodeChunk}
In this case, no past was cut, which means our threshold is too low. We can increase it by raising \code{alpha}, for example:
\begin{CodeChunk}
    \begin{CodeInput}
R> hdMTD_CUT(X, d = 40, S = c(1, 5, 10, 15, 17, 20, 27, 30, 35, 40),
+       alpha = 0.13)
    \end{CodeInput}
    \begin{CodeOutput}
[1] 1  5 27 35
    \end{CodeOutput}
\end{CodeChunk}
Just like our first use of the \code{hdMTD_BIC} function, we could have left the argument \code{S} empty and the function would set \code{S=1:d}. However, the CUT method verifies the relevance of each lag while \textbf{taking into account all other lags in \code{S}}. Not reducing our search to a subset \code{S} would make the function \code{hdMTD_CUT} work with sequences of size \code{d} (in this case, $40$) at all times, increasing significantly the running time. For example, as it is in the above example (with 10 elements in \code{S}), the function \code{hdMTD_CUT} took about 3 minutes to run. But if we add two more lags to S (for example \code{S = c(1,2,3,5,10,15,17,20,27,30,35,40)}), the process can take about 35 minutes.
Besides, the elements in \code{S} change the probabilities estimated within the function and might alter its output. In the example below, even though lags $-35$ and $-27$ were potential choices, when we reduce \code{S}, the output changes.
\begin{CodeChunk}
    \begin{CodeInput}
R> hdMTD_CUT(X, d = 40, S = c(1, 5, 17, 27, 30, 35), alpha = 0.13)
    \end{CodeInput}
    \begin{CodeOutput}
[1]  1  5 17 30
    \end{CodeOutput}
\end{CodeChunk}
Finally, the FSC estimator combines the FS and CUT estimators. It uses the function \code{hdMTD_FS} to reduce the set \code{1:d} to some subset \code{S} of size \code{l}, and then calls \code{hdMTD_CUT} to trim out the less important pasts. However, there is a distinction between employing the FS estimator followed by the CUT estimator and using the FSC estimator. In the FSC estimator, the sample is divided in half. The first half is used to compute the FS estimator, while the CUT estimator is computed from the second half. This is done to ensure that the CUT estimator is computed accordingly to the theory developed in \citet{Ost&Takahashi}.
\begin{CodeChunk}
\begin{CodeInput}
R> hdMTD_FSC(X, d = 40, l = 4, alpha = 0.1)
\end{CodeInput}
\begin{CodeOutput}
[1] 24 30
\end{CodeOutput}
\end{CodeChunk}
If you compare the output of the FS estimator with this output from FSC estimator, you might wonder about the appearance of the past $-24$ (since \code{hdMTD_FS(X, d = 40, l = 4)} returned \code{30, 15, 1, 27}). You need to take into account the splitting of the sample. In this example, the input set the FS estimator provided to FSC estimator was:
\begin{CodeChunk}
\begin{CodeInput}
R> hdMTD_FS(X[1:500], d = 40, l = 4)
\end{CodeInput}
\begin{CodeOutput}
[1] 11 30  7 24
\end{CodeOutput}
\end{CodeChunk}
Afterwards, the CUT estimator uses the second half of the sample to remove lags $-11$ and $-7$. To conclude, all arguments in the FSC estimator are exactly those used in FS and CUT estimators, and the order of the output no longer reflects the importance of each lag, as it did in FS. In fact, FS is the only estimator where the order of the output carries informative value.

Now let us assume we have estimated a set of relevant lags $\hat{\Lambda}$ from the data. If we wish to estimate the transition matrix \code{P} from the sample we can set \code{S} $=-\hat{\Lambda}$ and call the \code{empirical_probs} function.
\begin{CodeChunk}
    \begin{CodeInput}
R> head(empirical_probs(X, S = c(1, 15, 30)), 6)
    \end{CodeInput}
    \begin{CodeOutput}
  past_{ -30,-15,-1 } a p(a|past)
1                 000 0 0.5000000
2                 000 1 0.5000000
3                 001 0 0.3714286
4                 001 1 0.6285714
5                 010 0 0.6306306
6                 010 1 0.3693694
    \end{CodeOutput}
\end{CodeChunk}
\begin{CodeChunk}
    \begin{CodeInput}
R> empirical_probs(X, S = c(1, 15, 30), matrixform = TRUE)
    \end{CodeInput}
    \begin{CodeOutput}
            0         1
000 0.5000000 0.5000000
001 0.3714286 0.6285714
010 0.6306306 0.3693694
011 0.5065789 0.4934211
100 0.3861386 0.6138614
101 0.1987952 0.8012048
110 0.3888889 0.6111111
111 0.3697917 0.6302083
    \end{CodeOutput}
\end{CodeChunk}




When using the argument \code{matrixform = TRUE}, the output is displayed as a matrix. As a reading guide for transition matrices in the \pkg{hdMTD} package, rows represent past contexts (read from oldest to most recent). For example, the row \texttt{001}, in this case, gives the estimated transition probabilities given $x_{-30}=0$, $x_{-15}=0$, and $x_{-1}=1$ (as can be seen by comparing with the \code{data frame} output).

The package also features the function \code{oscillation}. This function returns the true \textbf{oscillations} for each lag if you provide it with an MTD \proglang{class} object.
\begin{CodeChunk}
    \begin{CodeInput}
R> oscillation(MTD)
    \end{CodeInput}
    \begin{CodeOutput}
     -1        -15        -30 
0.1233648 0.1017517 0.1846987 
    \end{CodeOutput}
\end{CodeChunk}
This function can also estimate the \textbf{oscillations} if you provide it with a sample of a MTD model and a set of lags $S$. In this case, the function will assume that \code{d = max(S)} and estimate the quantities in \eqref{eq:osc} for each lag in \code{S}.
\begin{CodeChunk}
    \begin{CodeInput}
R> oscillation(X, S = c(1, 15, 30))
    \end{CodeInput}
    \begin{CodeOutput}
       -1       -15       -30 
0.1076339 0.1166363 0.1675360
    \end{CodeOutput}
\end{CodeChunk}

The package also offers a method for estimating all MTD parameters using the expectation maximization algorithm (EM) via the \code{MTDest} function. This function was developed based on the algorithm proposed by \citet{EMforMTD} and produces an S3 object of class \code{MTDest}. It iteratively updates parameters to maximize log-likelihood until convergence or until a predetermined number of iterations is reached. To execute this procedure, an initial list of parameters must be provided through the \code{init} argument:
\begin{CodeChunk}
    \begin{CodeInput}
R> init <- list(
+       'lambdas' = c(0.01, 0.33, 0.33, 0.33),
+       'p0' = c(0.5, 0.5),
+       'pj' = rep(list(matrix(c(0.5, 0.5, 0.5, 0.5), ncol = 2, nrow = 2)), 3)
+       )
    \end{CodeInput}
    \end{CodeChunk}
This list must have the following entries: a vector named \code{lambdas} with the MTD weights; a list of matrices named \code{pj} containing the MTD past conditional distributions. If your MTD has an independent distribution the list should also have a vector called \code{p0} with the independent distribution. If no independent distribution is present, the first entry of \code{lambdas} must be set to zero.
The \code{MTDest} parameters update will halt by default if either the likelihood increase at some iteration becomes smaller than \code{M = 0.01} or if the number of iterations \code{nIter} reaches 100. Both these arguments can be modified by the user.

\begin{CodeChunk}
    \begin{CodeInput}
R> emMTD <- MTDest(X, S = c(1, 15, 30), init = init, iter = TRUE)
R> summary(emMTD)
    \end{CodeInput}
    \begin{CodeOutput}
Summary of EM estimation for MTD model:

Call:
MTDest(X = X, S = c(1, 15, 30), init = init, iter = TRUE)

Lags (-S): -1, -15, -30 
State space (A): 0, 1 

lambdas (weights):
       lam0       lam-1      lam-15      lam-30 
0.009353147 0.323151043 0.326490125 0.341005685 

Independent distribution p0:
    p0(0)     p0(1) 
0.3911887 0.6088113 

Transition matrices pj (one per lag):
 
 pj for lag j = -1:
          0         1
0 0.5982386 0.4017614
1 0.2641061 0.7358939
 
 pj for lag j = -15:
          0         1
0 0.1909446 0.8090554
1 0.5495934 0.4504066
 
 pj for lag j = -30:
          0         1
0 0.6978307 0.3021693
1 0.2113007 0.7886993

Log-likelihood: -623.879 

Iterations Report:
Number of updates: 9 
Last compared difference of logLik: 0.00700394 
\end{CodeOutput}
\end{CodeChunk}

In the output, \code{lambdas}, \code{pj}, and \code{p0} are the updated estimated parameters of the MTD model. Since \code{iter = TRUE}, the \code{summary()} also reports the iteration diagnostics, confirming convergence when the last log-likelihood difference dropped below \code{M = 0.01}. The fitted object additionally stores the log-likelihood of the model, which is displayed in the \code{summary()} output and can be directly accessed with the \code{logLik()} method. All available methods and accessors for \code{MTDest} objects are listed in the \code{print()} method output.

It is also possible to set \code{M} to \code{NULL}, and the function \code{MTDest} will only halt when the maximum number of iterations \code{nIter} is reached (by default \code{nIter = 100}). Additionally, there is the argument \code{oscillations} that can be set to \code{TRUE} if the user wishes the function to calculate the oscillations based on the outputted parameters.
\begin{CodeChunk}
    \begin{CodeInput}
R> emMTD <- MTDest(X, S = c(1, 15, 30), M = NULL, nIter = 9,
+       init = init, oscillations = TRUE)
R> summary(emMTD)
    \end{CodeInput}
    \begin{CodeOutput}
Summary of EM estimation for MTD model:

Call:
MTDest(X = X, S = c(1, 15, 30), M = NULL, init = init, nIter = 9, 
    oscillations = TRUE)

Lags (-S): -1, -15, -30 
State space (A): 0, 1 

lambdas (weights):
       lam0       lam-1      lam-15      lam-30 
0.009353147 0.323151043 0.326490125 0.341005685 

Independent distribution p0:
    p0(0)     p0(1) 
0.3911887 0.6088113 

Transition matrices pj (one per lag):
 
 pj for lag j = -1:
          0         1
0 0.5982386 0.4017614
1 0.2641061 0.7358939
 
 pj for lag j = -15:
          0         1
0 0.1909446 0.8090554
1 0.5495934 0.4504066
 
 pj for lag j = -30:
          0         1
0 0.6978307 0.3021693
1 0.2113007 0.7886993

Log-likelihood: -623.879 

Oscillations:
       -1       -15       -30 
0.1079753 0.1170953 0.1659095
\end{CodeOutput}
\end{CodeChunk}
In this example, the function \code{MTDest} stopped after $9$ updates because we have set \code{nIter = 9}. The estimated parameters in the output should be exactly the same as before, as there were the same number of iterations.

Since \code{MTDest} objects do not carry a global transition matrix the \code{transitP()} accessor is not available for it. However, the user can coerce an \code{MTDest} object to an \code{MTD} via the \code{as.MTD()} function and thus access any methods for it:
\begin{CodeChunk}
\begin{CodeInput}
R> emMTD <- MTDest(X, S = c(1, 15, 30), init = init)
R> MTD_hat <- as.MTD(emMTD)
R> transitP(MTD_hat)
\end{CodeInput}
\begin{CodeOutput}
            0         1
000 0.4972860 0.5027140
001 0.3893107 0.6106893
010 0.6143813 0.3856187
011 0.5064060 0.4935940
100 0.3313766 0.6686234
101 0.2234013 0.7765987
110 0.4484718 0.5515282
111 0.3404966 0.6595034
\end{CodeOutput}
\end{CodeChunk}

\subsection[Testing hdMTD]{Testing \pkg{hdMTD}}

\subsubsection{Simulated Data.}

To assess the precision of our estimators, we perform the following experiment: given a fixed MTD model with state space $\cA=\{0,1\}$ and set of relevant lags $\Lambda=\{-5,-1\}$,
\begin{CodeChunk}
\begin{CodeInput}
R> set.seed(123)
R> Lambda <- c(1, 5)
R> A <- c(0, 1)
R> lam0 <- 0.01
R> p0 <- c(0.5, 0.5)
R> MTD <- MTDmodel(Lambda, A, lam0, p0 = p0)
\end{CodeInput}
\end{CodeChunk}
we generate $N_{rep}=100$ replications of size \code{N = 10000} using the \code{perfectSample} function.

For each of the $N_{rep}$ replications and for values of $m\in\{1000,1500,2000,2500,3000,5000,10000\}$, we apply the \code{hdMTD} function on the first $m$ elements of the replication with \code{method = FS} and parameters \code{d = 100} and \code{l = 2}, obtaining an output \code{S} denoted $S_{j,m}$. Here, the variable $1\leq j\leq N_{rep}$ indicates which replication was used. 
Subsequently, we estimate $\hat{\Prob}_m(\cdot|00_{S_{j,m}})$ \footnote{For instance, if $S_{j,m}=\{-56,-3\}$, then $\hat{\Prob}_m(\cdot|00_{S_{j,m}})$ represents the estimated conditional probabilities given that $X_{-56}=0$ and $X_{-3}=0$. Although any pasts in $\mathcal{A}^2$ could have been chosen, we opted for past $00$ for simplicity.} and compute:
$$\Delta_{FS}(j,m) = d_{TV}(\hat{\Prob}_m(\cdot|00_{S_{j,m}}),\Prob(\cdot|00_{\Lambda}))=|\hat{\Prob}_m(0|00_{S_{j,m}})-\Prob(0|00_{\Lambda})|\;,$$
and
$$\Delta_{FS}^{std}(j,m) = \dfrac{|\hat{\Prob}_m(0|00_{S_{j,m}})-\Prob(0|00_{\Lambda})|}{\min\{\Prob(0|00_{\Lambda}),\Prob(1|00_{\Lambda})\}}\;,$$
for the different values of $j$ and $m$.

Afterwards, we computed two additional estimators for comparison. The first one, termed \textit{Naive} estimator, utilizes pasts made of the 5 most recent past states.
\footnote{We reduced the order from $100$ to $5$ when calculating the \textit{Naive} estimator because the probability of encountering a sequence of $100$ zeros in these realizations is exceedingly low.} We denote $\Delta_{Naive,5}(j,m)$ and $\Delta_{Naive,5}^{std}(j,m)$ the quantities $\Delta_{FS}(j,m)$ and $\Delta_{FS}^{std}(j,m)$, respectively, computed with $\hat{\Prob}_m(0|00_{S_{j,m}})$ replaced by $\hat{\Prob}_m(0|0\dots 0_{\lk -5,-1 \rk})$. Since the true order of the MTD is $5$, the \textit{Naive} estimator is consistent. The second estimator, called \textit{Oracle of size 2}, represents the best possible estimator using any two lags within the set $\lk -100,-1\rk$. The \textit{Oracle of size 2} is defined as $S_{j,m}^*= \underset{|S|=2 : S\subset \lk -100,-1\rk}{\arg\min} d_{TV}(\hat{\Prob}_m(\cdot|00_S),\Prob(\cdot|00_{\Lambda}))$. 
We denote $\Delta_{Oracle}(j,m)$ and $\Delta_{Oracle}^{std}(j,m)$ the quantities $\Delta_{FS}(j,m)$ and $\Delta_{FS}^{std}(j,m)$, respectively, computed with $\hat{\Prob}_m(0|00_{S^*_{j,m}})$ in the place of $\hat{\Prob}_m(0|00_{S_{j,m}})$.
Note that \textit{Oracle of size 2} cannot be computed without prior knowledge of the true MTD parameters. Here, we use it as a benchmark to evaluate the performance of the FS and Naive estimators.

\begin{center}
\begin{figure}[h!]
\centering
\includegraphics[width=1\textwidth]{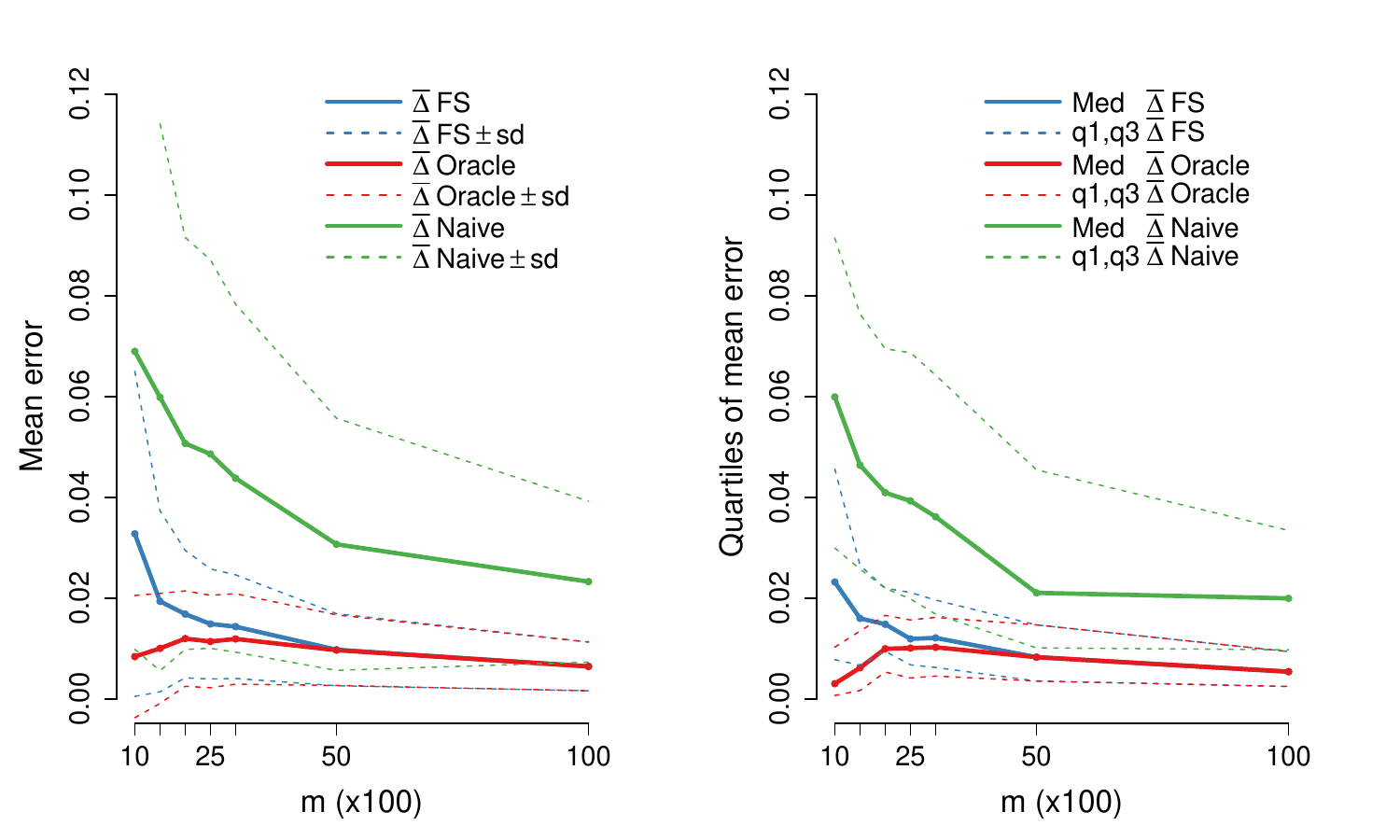}
\caption{Estimators mean error across $N_{rep}=100$ replications.} \label{fig:grafx}
\end{figure}
\end{center}

Table \ref{tab:tabelax} and the left plot in Figure \ref{fig:grafx} display the averages of the total variation distances (e.g., $\bar{\Delta}_{FS}(m)= \displaystyle\sum_{j=1}^{N_{rep}} \frac{\Delta_{FS}(j,m)}{N_{rep}}$) through the $N_{rep}=100$ replications, and their behavior when increasing the value of $m$. Figure \ref{fig:grafx} left plot also depicts, with dashed lines, the intervals generated by adding and subtracting the standard deviation errors. The right plot in Figure \ref{fig:grafx} show the evolution of the quartiles of the computed distances (q1: \textit{1-th quartile}, Med: \textit{median}, and q3: \textit{3-rd quartile}).

\begin{table}[h!]
\centering
\begin{tabular}{lccccccc}
\hline
$m$ & 1000 & 1500 & 2000 & 2500 & 3000 & 5000 & 10000 \\
\hline
$\bar{\Delta}_{\mathrm{FS}}(m)$           & 0.03279 & 0.01940 & 0.01686 & 0.01492 & 0.01438 & 0.00980 & 0.00649 \\
$\bar{\Delta}_{\mathrm{Oracle}}(m)$       & 0.00844 & 0.01006 & 0.01201 & 0.01143 & 0.01194 & 0.00971 & 0.00649 \\
$\bar{\Delta}_{\mathrm{Naive},5}(m)$      & 0.06898 & 0.05988 & 0.05072 & 0.04862 & 0.04381 & 0.03072 & 0.02330 \\
\hline
$\bar{\Delta}^{\mathrm{std}}_{\mathrm{FS}}(m)$       & 0.07452 & 0.04410 & 0.03832 & 0.03390 & 0.03268 & 0.02228 & 0.01475 \\
$\bar{\Delta}^{\mathrm{std}}_{\mathrm{Oracle}}(m)$   & 0.01917 & 0.02287 & 0.02729 & 0.02598 & 0.02714 & 0.02207 & 0.01475 \\
$\bar{\Delta}^{\mathrm{std}}_{\mathrm{Naive},5}(m)$  & 0.15676 & 0.13608 & 0.11525 & 0.11049 & 0.09957 & 0.06980 & 0.05296 \\
\hline
\end{tabular}
\caption{\label{tab:tabelax} Mean error of estimators ($d=5$).}
\end{table}

As can be seen, the mean error of the FS estimator seems to converge to that of the \textit{Oracle of size 2} estimator as $m$ increases. In fact, when using the full realizations (i.e., $m=10000$), the FS method and the \textit{Oracle of size 2} agreed that the relevant lag set was $\{-5,-1\}$ across all $100$ replications. For $m=5000$, they only disagreed in 3 replications. This agreement reflects the high precision and consistency of the FS estimator as $m$ grows.
In contrast, the Naive estimator —which has access to the true relevant lags — exhibited substantially higher errors in total variation distance to $\Prob(\cdot | 00_\Lambda)$ across all values of $m$. This suggests that simply including the correct lags is not sufficient: without appropriate model selection the estimator may fail to capture the true conditional distribution effectively, especially for smaller sample sizes.

\subsection{Analysis of real-world data.}

\subsubsection{Temperatures in Brazil.} 

The \code{hdMTD} package includes the \code{tempdata} dataset, a \proglang{tibble} containing hourly temperature measurements from Brasília, Brazil. The dataset comprises three columns: \code{DATE} (ranging from January 01, 2003 to August 31, 2024), \code{TIME} (hourly intervals from 00:00 to 23:00 in UTC-3), and \code{MAXTEMP} (maximum temperature in Celsius recorded within each hour interval). With 189,936 observations spanning 7,914 complete days, this dataset provides comprehensive coverage of Brasília's thermal patterns at 15.78°S, 47.92°W (altitude: 1,159.54m). The data was collected by Brazil's National Institute of Meteorology \citep{inmet2024} and can be loaded in \proglang{R} using \code{data("tempdata")}.

The dataset contains 3,035 missing temperature values (i.e., not applicable or \code{NA} data). Preliminary analysis showed that $95\%$ of these values occurred before August 5, 2010. Therefore, for this analysis, we truncated the dataset to include only observations from this date onward, resulting in 123,384 hourly measurements (5,141 days) and 155 remaining \code{NA} values. In the remaining data, short sequences of missing values (up to 6 consecutive \code{NA}s) were imputed using the mean of the most recent non-missing value and the next available non-missing value within six hours. For longer sequences (7 or more consecutive \code{NA}s), each missing value was replaced sequentially by the mean of up to three values: the temperature at the same hour on the previous day, the previous hour, and the next hour (ignoring \code{NA}s in this calculation).

We then aggregated the data in a new dataset (\code{temp}) by computing the daily mean of the hourly maximum temperatures (\code{MAXTEMP}), producing a time series of daily maximum temperatures.
    \begin{CodeChunk}
    \begin{CodeInput}
R> head(temp, 4)   
    \end{CodeInput}
    \begin{CodeOutput}
# A tibble: 4 × 2
  DATE       MAXTEMP
  <date>       <dbl>
1 2010-08-05    20.7
2 2010-08-06    20.5
3 2010-08-07    21.8
4 2010-08-08    22.3
    \end{CodeOutput}
    \end{CodeChunk}

\begin{center}
\begin{figure}[ht!]
\centering
    \includegraphics[width=0.8\linewidth]{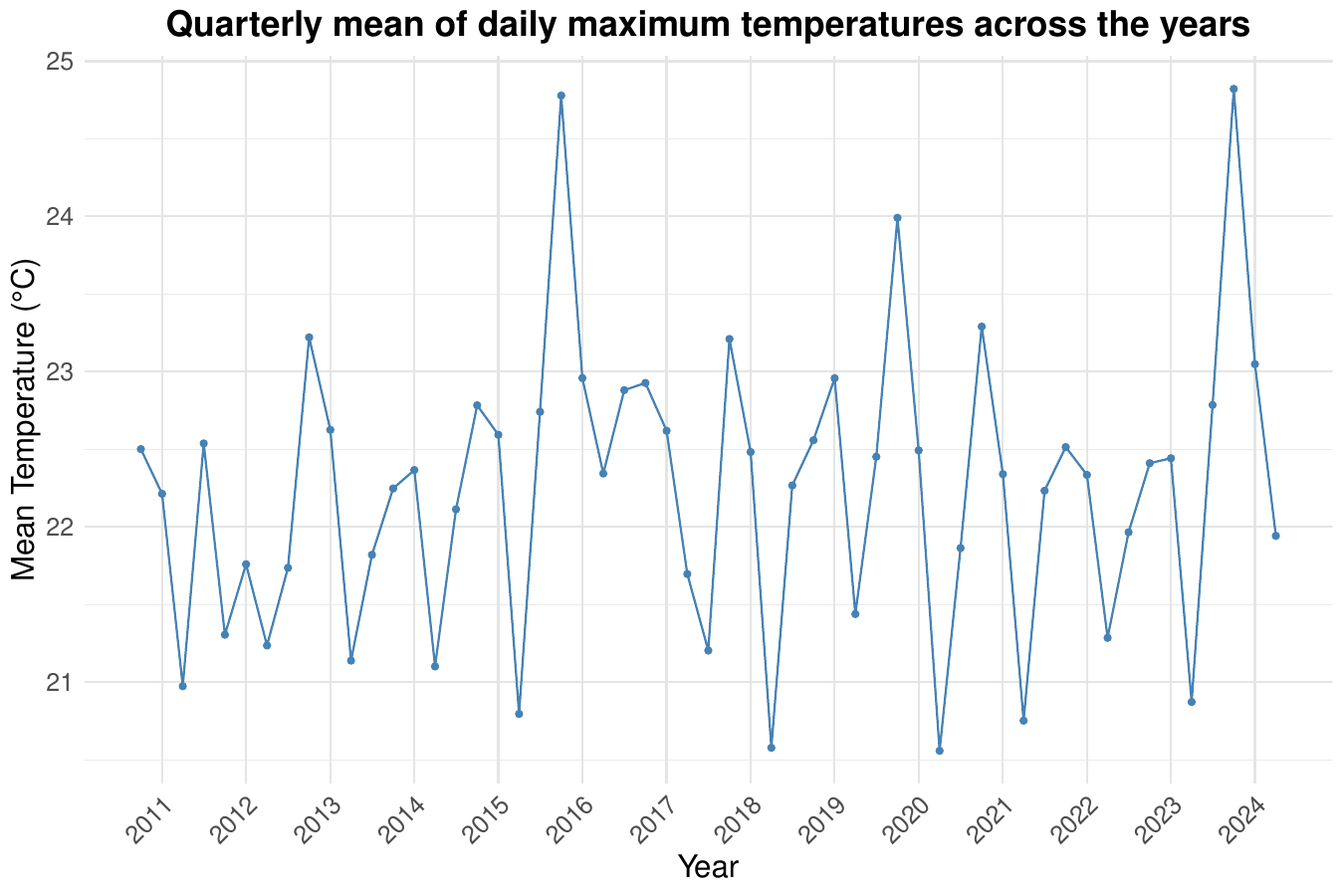}
    \caption{Time series with quarterly mean of daily maximum temperatures in Brasília, Brazil.}
    \label{fig:temp_serie}
\end{figure}
\end{center}

Figure~\ref{fig:temp_serie} shows the quarterly means of the daily maximum temperatures from 2010 onward. Each point represents the average of all daily maxima within the corresponding trimester.

Finally, we discretized the continuous temperature data into two categories of equal range. This partitioning creates a variable \code{MAXTEMP1} with two states representing distinct thermal regimes.
\begin{CodeChunk}
\begin{CodeInput}
R> head(temp, 4)   
\end{CodeInput}
\begin{CodeOutput}
# A tibble: 4 × 3
  DATE       MAXTEMP MAXTEMP1
  <date>       <dbl>    <dbl>
1 2010-08-05    20.7        1
2 2010-08-06    20.5        1
3 2010-08-07    21.8        2
4 2010-08-08    22.3        2
    \end{CodeOutput}
    \end{CodeChunk}
The categorization scheme is as follows:
\begin{itemize}
\item \textbf{Category 1}: Temperatures in [12.046,20.923)°C (coded as 1), accounting for $26.1\%$ of the data.
\item \textbf{Category 2}: Temperatures in [20.923,29.8]°C (coded as 2), accounting for $73.9\%$ of the data.
\end{itemize}
With this setting, we can use the package to determine the set of relevant pasts for predicting temperature regimes.  We must first reverse the dataset since \pkg{hdMTD} package functions assume that the sample is sorted from the latest observation to the oldest, and then we can apply the FS method for trying to capture very long past dependencies.

\begin{CodeChunk}
\begin{CodeInput}
R> Temp12 <- rev(temp$MAXTEMP1)
R> hdMTD_FS(Temp12, d = 400, l = 3)
\end{CodeInput}
\begin{CodeOutput}
[1] 1 364 6
\end{CodeOutput}
\end{CodeChunk}

These results indicate that, when modeling the regime transitions between high and low temperatures, the current state is strongly influenced by the previous day, as well as by lags of $-364$ and $-6$ days. This suggests the presence of both annual and weekly cycles in the behavior of daily maximum temperature regimes. The output indicates $-364$ as the maximum lag to consider. Therefore rerunning \code{hdMTD_FS} with \code{d = 364} is recommended since the computations within the FS algorithm will use a more robust estimator for the required transition probabilities. In this case, the exact same output for the relevant lags is obtained.

Next, we assess the improvement in predictive performance provided by our lag selection methods. To do this, we split the dataset into training and testing subsets. We reserve the last year of available data, from September 1, 2023 to August 31, 2024 (a total of $366$ days), for testing. The remaining $4775$ days are used for training.

\begin{CodeChunk}
\begin{CodeInput}
R> ndays <- nrow(temp %>% filter(DATE >= "2023-09-01"))
R> Temp12_Train <- Temp12[-seq_len(ndays)]
R> Temp12_Test <- Temp12[seq_len(ndays)]
\end{CodeInput}
\end{CodeChunk}
We rerun the \code{hdMTD_FS} function on \code{Temp12_Train} to account for the reduced sample size. Interestingly, the resulting set of selected lags remains unchanged ($-1$, $-364$, and $-6$), indicating that these dependencies are stable and persist even when the last year of data is excluded.

\begin{CodeChunk}
\begin{CodeInput}
R> hdMTD_FS(Temp12_Train, d = 364, l = 3)
\end{CodeInput}
\begin{CodeOutput}
[1] 1 364 6
\end{CodeOutput}
\end{CodeChunk}

Running the FS method with \code{l = 3} constrains the output to a set of three lags. While these lags are the most significant identified by the algorithm, there may be other relevant past observations, or some of the selected lags might have limited actual impact. We choose \code{l = 3} as it offers a good trade-off between informativeness and computational efficiency.
 If the true relevant lag set $\Lambda$ satisfies $|\Lambda| < 3$, the FS output certainly contains potentially irrelevant pasts. The \pkg{hdMTD} package provides a few strategies to refine the FS output and eliminate spurious lags:
 
1. Apply the CUT algorithm to the FS output:
\begin{CodeChunk}
\begin{CodeInput}
R> hdMTD_CUT(Temp12_Train, d = 364, S = c(1, 364, 6))
\end{CodeInput}
\begin{CodeOutput}
[1]   1   6 364
\end{CodeOutput}
\end{CodeChunk}
In this example, the CUT method retained all three lags, indicating that it considers all elements in \code{S = c(1, 6, 364)} to be relevant.

2. Use BIC-based model selection on the FS output:
\begin{CodeChunk}
\begin{CodeInput}
R> hdMTD_BIC(Temp12_Train, d = 364, S = c(1, 364, 6), minl = 1, maxl = 3,
+       byl = TRUE, BICvalue = TRUE)
\end{CodeInput}
\begin{CodeOutput}
       1             1,364           1,6,364    smallest: 1,6,364 
1720.801          1690.543          1674.080             1674.080
\end{CodeOutput}
\end{CodeChunk}
This performs an exhaustive search over the $2^3 - 1 = 7$ possible non-empty subsets of \code{S = c(1, 6, 364)}, selecting the configuration with the lowest Bayesian information criterion (BIC). Since \code{byl = TRUE}, the output reports the best-performing subset (top row) and its corresponding BIC value (bottom row) for each possible subset size (from 1 to 3 lags). In this example, the full set \code{c(1, 6, 364)} yields the lowest BIC, and is thus selected as the optimal lag configuration.

The package also includes a method, \code{FSC}, which automatically runs FS followed by CUT using the FS output. As discussed in Section~\ref{FSsec}, the consistent application of this procedure involves sample splitting, therefore the FSC method uses the first half of the data for FS, and the second half for CUT.
Due to the sample split, the FS step in FSC has access to only half the observations. Despite this reduction, in this example, it still selected the same lag set as before. The subsequent CUT step confirmed all selected lags as relevant, so none were removed.

\begin{CodeChunk}
\begin{CodeInput}
R> hdMTD_FSC(Temp12_Train, d = 364, l = 3)
\end{CodeInput}
\begin{CodeOutput}
[1]   1   6 364
\end{CodeOutput}
\end{CodeChunk}

 All methods suggest that the set \code{S = c(1, 6, 364)} is more appropriate for model specification than any of its subsets.
 It is possible that a model with a larger set of lags could result in an even better choice, but increasing the number of lags would substantially raise the computational cost, as it would require re-running FS with a larger \code{l} and repeating the refinement steps.
 Thus, for the purposes of this example, we proceed by estimating the transition probabilities of an MTD model using the selected set \code{S = c(1, 6, 364)}.

\begin{CodeChunk}
\begin{CodeInput}
R> P_FS <- empirical_probs(Temp12_Train, S = c(1, 6, 364), matrixform = T)
R> P_FS
\end{CodeInput}
\begin{CodeOutput}
             1         2
111 0.86626140 0.1337386
112 0.24736842 0.7526316
121 0.77157360 0.2284264
122 0.13318777 0.8668122
211 0.78846154 0.2115385
212 0.10972569 0.8902743
221 0.57506361 0.4249364
222 0.07283555 0.9271645
\end{CodeOutput}
\end{CodeChunk}
Recall that the symbols in the rows represent occurrences at the pasts in $S$, ordered from oldest to most recent. For example sequence $112$ (in the second row) means that symbol $1$ occurred at lag $-364$, symbol $1$ at lag $-6$ and symbol $2$ at lag $-1$.

To evaluate whether our lag selection method improves estimation, we compare it with a classical Markov chain selection method. The most traditional approach estimates the order of the Markov chain by minimizing the Bayesian information criterion (BIC), without assuming the MTD structure. We then compute the penalized log-likelihood for classical Markov chains of orders $1$ through $6$. 

Note that the number of parameters in a full Markov chain model increases exponentially with the order, with the BIC penalty term given by $\frac{\log(n) |\mathcal{A}|^d(\mathcal{A}-1)}{2}$. As such, higher-order models are heavily penalized, and large orders are unlikely to be selected. We limit the analysis to order $6$ because the forward selection (FS) method identified lag $6$ as potentially relevant.

To ensure comparability of log-likelihoods across models of different orders, we fix \code{d = 6}. Using the \code{countsTab} function, we count the occurrences of all sequences of length $d+1$:

\begin{CodeChunk}
\begin{CodeInput}
R> ct <- countsTab(Temp12_Train, d = 6) 
R> head(ct, 4)
\end{CodeInput}
\begin{CodeOutput}
# A tibble: 4 × 8
     x6    x5    x4    x3    x2    x1     a   Nxa
  <dbl> <dbl> <dbl> <dbl> <dbl> <dbl> <dbl> <int>
1     1     1     1     1     1     1     1   342
2     1     1     1     1     1     1     2    62
3     1     1     1     1     1     2     1    18
4     1     1     1     1     1     2     2    62
\end{CodeOutput}
\end{CodeChunk}

Next, we compute the transition probabilities and log-likelihoods for Markov chains of increasing order: \textbf{MC1} (with \code{S = 1}), \textbf{MC2} (\code{S = c(1, 2)}), up to \textbf{MC6} (\code{S = c(1, 2, 3, 4, 5, 6)}). For instance, for the third-order Markov chain (\textbf{MC3}), we use the \code{freqTab} function to estimate transition probabilities:

\begin{CodeChunk}
\begin{CodeInput}
R> ft <- freqTab(S = c(1, 2, 3), A = c(1, 2), countsTab = ct) 
R> head(ft, 4)
\end{CodeInput}
\begin{CodeOutput}
# A tibble: 4 × 7
     x3    x2    x1     a Nxa_Sj Nx_Sj qax_Sj
  <dbl> <dbl> <dbl> <dbl>  <int> <int>  <dbl>
1     1     1     1     1    587   726  0.809
2     1     1     1     2    139   726  0.191
3     1     1     2     1     38   206  0.184
4     1     1     2     2    168   206  0.816
\end{CodeOutput}
\end{CodeChunk}

We then compute the log-likelihood and the corresponding BIC value:

\begin{CodeChunk}
\begin{CodeInput}
R> LL <- sum(log(ft$qax_Sj) * ft$Nxa_Sj)
R> freeParam <- 2^3 * 1
R> BICMC3 <- -LL + 0.5 * log(length(Temp12_Train)) * freeParam
R> BICMC3
\end{CodeInput}
\begin{CodeOutput}
[1] 1854.029
\end{CodeOutput}
\end{CodeChunk}

\begin{table}[ht!]
\centering
\begin{tabular}{lcccccc}
\hline
 & MC1 & MC2 & MC3 & MC4 & MC5 & MC6 \\
\hline
BIC & 1869.162 & \textbf{1850.598} & 1854.029 & 1877.888 & 1925.962 & 2031.679 \\
\hline
\end{tabular}
\caption{BIC values computed for classical Markov chain models of different orders.}
\label{tab:BIC_MC_orders}
\end{table}

Table \ref{tab:BIC_MC_orders} presents the BIC values obtained for classical Markov chain models of orders $1$ through $6$. These values reflect the trade-off between model complexity and goodness of fit. As expected, the BIC increases for higher-order models due to the exponential growth in the number of parameters, which leads to stronger penalization. The model with the lowest BIC is \textbf{MC2}, corresponding to a second-order Markov chain with $S = \{1, 2\}$. Therefore, using the classical model selection strategy based on BIC minimization, this would be the preferred model for the data. The corresponding empirical transition matrix is:

\begin{CodeChunk}
\begin{CodeInput}
R> P_MC2 <- empirical_probs(Temp12_Train, S = c(1, 2), matrixform = TRUE)
R> P_MC2
\end{CodeInput}
\begin{CodeOutput}
            1         2
11 0.77813505 0.2218650
12 0.16326531 0.8367347
21 0.60233918 0.3976608
22 0.09064976 0.9093502
\end{CodeOutput}
\end{CodeChunk}
We now compare the predictive capabilities of this classical selection method with our proposed method (\textbf{FS}). Additionally, we include a naive baseline model (\textbf{Ind}), which assumes independence between observations: each symbol is drawn according to the overall empirical distribution of the training set.

\begin{CodeChunk}
\begin{CodeInput}
R> P_Ind <- prop.table(table(Temp12_Train))
R> P_Ind
\end{CodeInput}
\begin{CodeOutput}
Temp12_Train
        1         2 
0.2672251 0.7327749
\end{CodeOutput}
\end{CodeChunk}

This output displays the relative frequency of each temperature regime in the training sample. \textit{State 1} accounts for $26.7\%$ and \textit{State 2} for $73.3\%$ of the training data.

To quantitatively evaluate predictive performance, we performed a Monte Carlo simulation where each model was used to generate predictions for the $366$-day test set, repeated $1000$ times to account for variability introduced by random sampling. In each repetition, predictions were generated using: the empirical distribution \code{P_Ind} for the independent model (\textbf{Ind}), the second-order transition matrix \code{P_MC2} for the classical model (\textbf{MC2}), and the sparse high-order matrix \code{P_FS} from our proposed selection model (\textbf{FS}).

For the independent model (\textbf{Ind}), we simply sample $366$ symbols from the distribution \code{P_Ind}. For the \textbf{FS} and \textbf{MC2} models, predictions must be conditioned on previous observations. Specifically, we use the observed test and train sequences to extract the necessary histories and determine the next-step prediction.

Given that low-temperature days (\textit{State 1}) are relatively rare, especially in the test data - only $64$ out of $366$ days, about $17.5\%$, compared to $26.7\%$ in the training set - we paid particular attention to the performance of the models in predicting these events.

For each repetition, we computed standard performance metrics for binary classification (where \textit{State 1} represents the target regime and \textit{State 2} the alternative). Table \ref{tab:metrics} summarizes the mean results across 1000 repetitions.

\begin{table}[ht!]
\centering
\begin{tabular}{lcccc}
\hline
Metric & Formula & Ind (\%) & MC2 (\%) & FS (\%) \\
\hline
Accuracy              & (TP + TN) / (TP + TN + FP + FN)       & 65.09 & 82.37 & 83.49 \\
Precision (PPV)       & TP / (TP + FP)                        & 17.46 & 49.81 & 52.49 \\
Sensitivity (Recall)  & TP / (TP + FN)                        & 26.72 & 58.70 & 62.95 \\
Specificity           & TN / (TN + FP)                        & 73.22 & 87.38 & 87.85 \\
F1-score              & 2(PPV × Recall) / (PPV + Recall)      & 21.12 & 53.89 & 57.24 \\
\hline
\end{tabular}
\caption{Model performance metrics. Values represent means across 1000 replications.
Models: independent (\textbf{Ind}), second-order Markov chain (\textbf{MC2}), and forward stepwise (\textbf{FS}). 
TP = true positive; TN = true negative; FP = false positive; FN = false negative.}
\label{tab:metrics} 
\end{table}

\begin{figure}[ht!]
  \centering
    \includegraphics[width=0.9\linewidth]{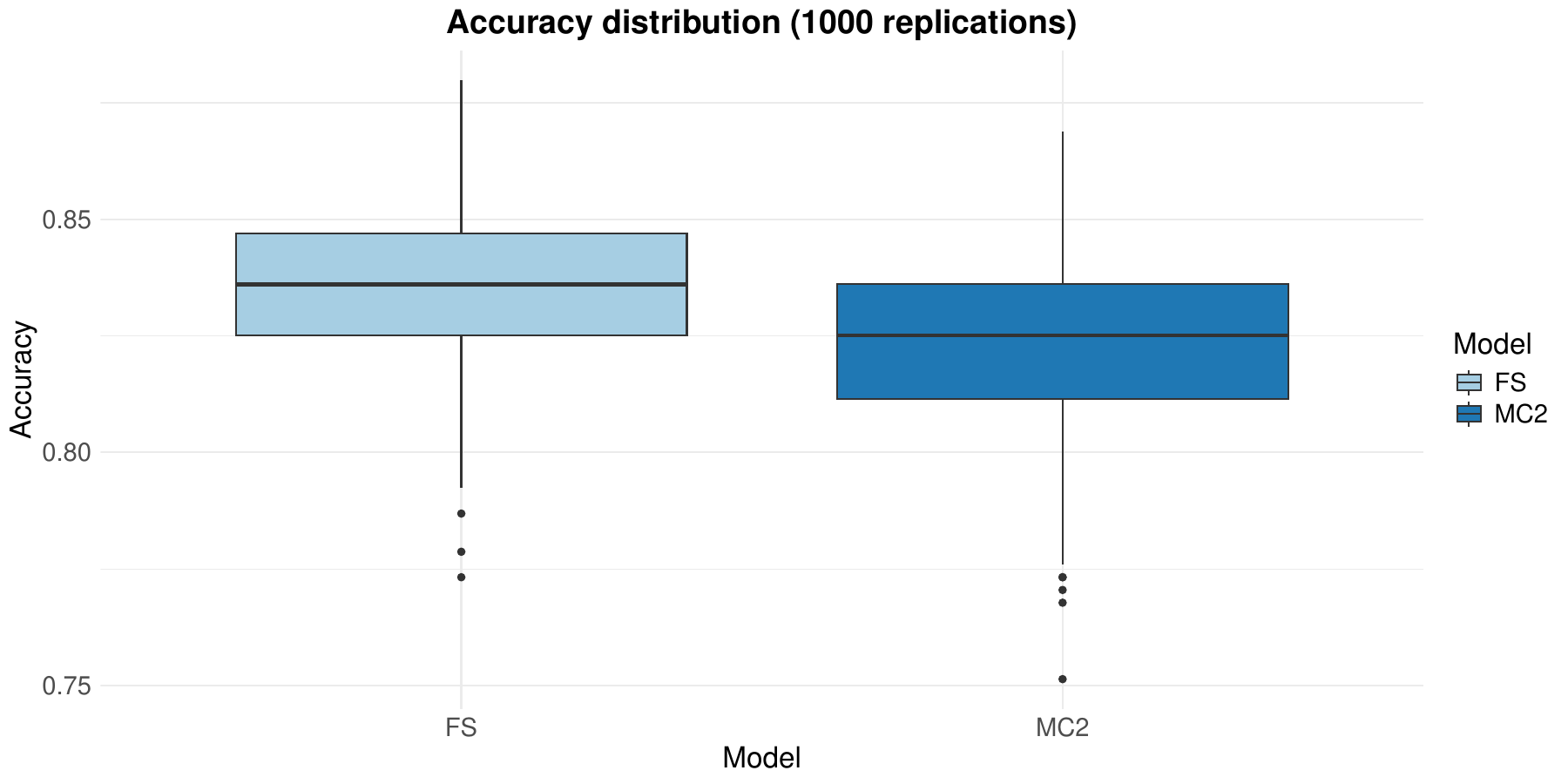}
    \caption{Exploratory analysis of accuracies along $1000$ replications for models \textbf{FS} and \textbf{MC2}.}
    \label{fig:acc}
\end{figure}

The simulation results (Table~\ref{tab:metrics} and Figure~\ref{fig:acc}) demonstrate \textbf{FS's consistent outperformance} across all evaluation metrics:

\begin{itemize}
\item \textbf{Accuracy Dominance}: \textbf{FS} achieves the highest prediction accuracy ($83.5\%$), showing an improvement over \textbf{MC2} ($82.4\%$) and the Independent model ($65.1\%$). Figure \ref{fig:acc} displays the accuracy distribution for both models among the $1000$ replications indicating higher quartile values for \textbf{FS}.
\item \textbf{Sensitivity}: \textbf{FS} correctly identifies $63\%$ of lower temperature days, a \textbf{$7.3\%$ increase} over \textbf{MC2} ($58.7\%$) and \textbf{$2.4$ times better} than the Independent model ($26.7\%$)
\item \textbf{Precision-Recall Balance}:
\textbf{FS} maintains the best F1-score ($57.2\%$), balancing precision ($52.5\%$) and recall ($63\%$). While \textbf{MC2} shows comparable precision ($49.8\%$), its lower recall demonstrates \textbf{FS}'s advantage in capturing rare \textit{State 1} events
\end{itemize}

As discussed in Section \ref{FSsec}, the FS method selects lags sequentially based on the empirical quantities $\hat{\nu}_{n,j,S}$ defined in \eqref{eq:hatnu}. These quantify the additional predictive power that lag $j$ contributes beyond the information in $S$, as estimated from the sample. To better understand the algorithm, we compute the values of $\hat{\nu}_{n,j,S}$ in a step-by-step fashion. At each step, we estimate $\hat{\nu}_{n,j,S}$ for all candidate lags $j \in \{ -364, \dots, -1\}$ not yet in the active set $S$, and select the lag with the largest value. Initially, $S=\emptyset$, and after each selection, the chosen lag is added to $S$ for the next iteration. This mimics the core logic of the FS method. 

\begin{figure}[ht!]
  \centering
\includegraphics[width=1\linewidth]{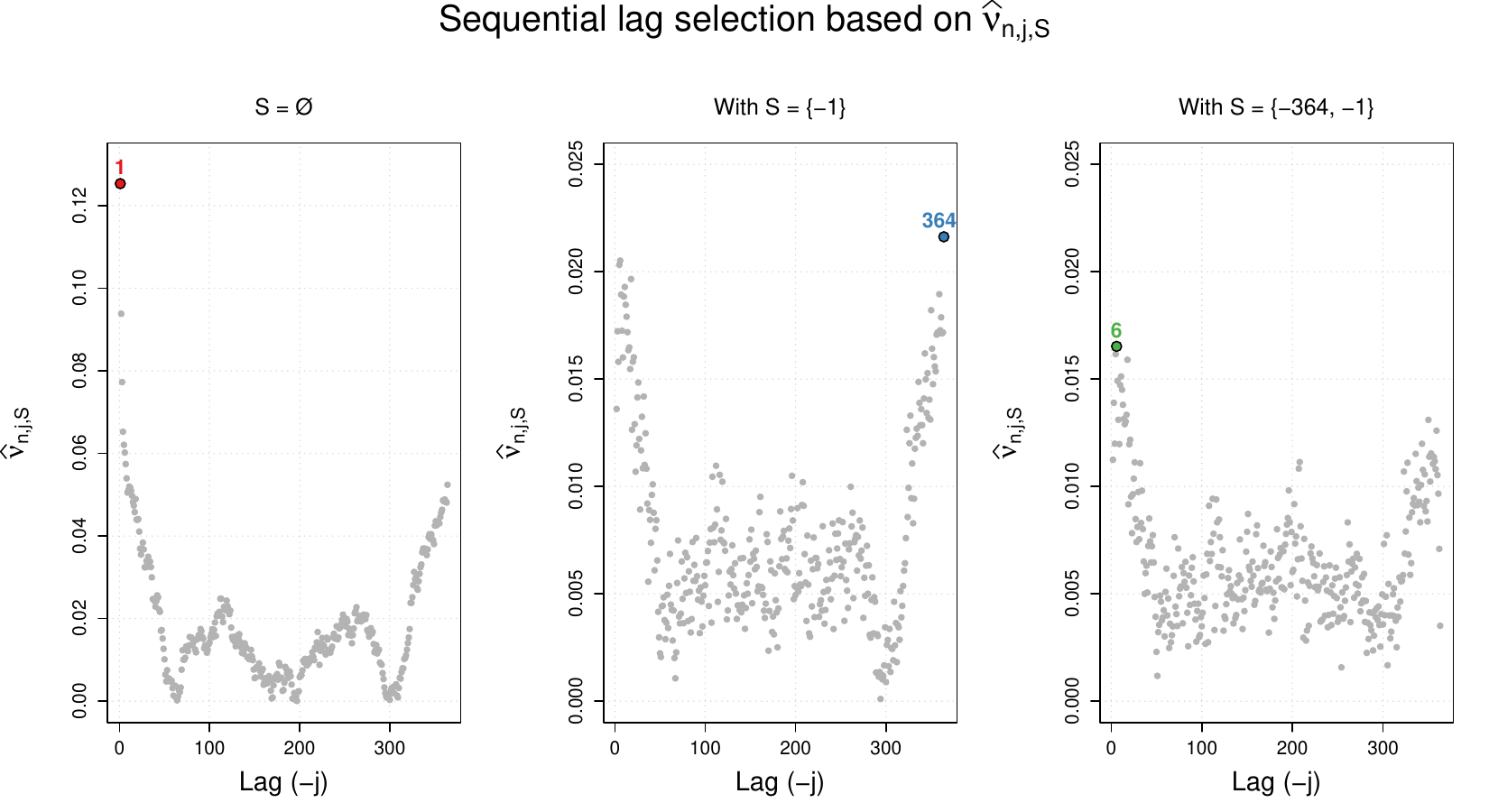}
  \caption{FS sequential step analysis through $\hat{\nu}_{n,j,S}$. Note that the scale of the first plot is different from the remaining two.}
    \label{fig:nu_plot}
\end{figure}

Figure \ref{fig:nu_plot} summarizes the results of this analysis in three graphics. The left plot shows the values of $\hat{\nu}_{n,j,S}$ for $S=\emptyset$ and each possible value $-j$ from $1$ to $364$, where lag $-1$ stands out with the highest value. The center plot shows the values $\hat{\nu}_{n,j,S}$ for each $-j \in \{2,\dots, 364\}$ when calculated jointly with lag $-1$ (i.e., $S = \{-1\}$). In this case, the lag  $-364$ attains the maximum of $\hat{\nu}_{n,j,S}$. Finally, the right plot displays the values of $\hat{\nu}_{n,j,S}$ for each lag $-j \in \{2,\dots, 363\}$ calculated given the information in lags $-1$ and $-364$ (i.e., $S = \{-364, -1\}$) identifying lag $-6$ as the most significant addition in this step as expected.
\newpage

\begin{CodeChunk}
    \begin{CodeOutput}
=== Final Selection Results ===
  Step Selected_Lag     nu
1    1            1 0.12533879
2    2          364 0.02162680
3    3            6 0.01652459
    \end{CodeOutput}
\end{CodeChunk}

\section*{Computational details}
\proglang{R} and all associated packages are publicly available via the Comprehensive \textsf{R} Archive Network (CRAN) at \url{https://CRAN.R-project.org/}. All analyses in this study were performed using \textsf{R}~version~4.4.1 \citep{R-base} with the \pkg{hdMTD} package (v0.1.3), along with the following auxiliary packages: \pkg{dplyr} \citep{pkg:dplyr}, \pkg{future} \citep{future}, \pkg{future.apply} \citep{future}, \pkg{ggplot2} \citep{pkg:ggplot2}, \pkg{lubridate} \citep{pkg:lubridate}, \pkg{purrr} \citep{pkg:purrr}, and \pkg{tidyr} \citep{pkg:tidyr}.
\noindent

The complete package source code, data, and supporting materials are available at CRAN and in the GitHub repository: \url{https://github.com/MaiaraGripp/hdMTD}. 
Additionally, the R scripts used to reproduce the results presented in Section \ref{sec:usingMTD} are available in the subdirectory \texttt{article-demos} of the same repository: \url{https://github.com/MaiaraGripp/hdMTD/tree/master/article-demos}. 
These demonstration scripts are not included in the installed version of the package.


\section{Conclusions} \label{sec:summary}

In this article, we present an \proglang{R} package called \pkg{hdMTD} 
for non-parametric estimation of high-dimensional MTD models, a sub-class of high-order Markov chains. We illustrate through analysis of simulated and empirical data application that the \pkg{hdMTD} package is particularly useful in context of statistical analysis of categorical time series with long-range dependencies.

Given a sample from an MTD model, \pkg{hdMTD} can retrieve the set of relevant pasts (lags) even when the order of the model is proportional to the sample size. Our implementation of the lag selection method can be seen as a feature selection method for non-parametric categorical time series and fills an important gap in the literature. The key feature of our lag selection method is that we can estimate the set of candidates for relevant lags a priori without the need to estimate the $d$-dimensional joint probability measure of a Markov chain of order $d$. Once the set of candidates is estimated, we apply an adaptive thresholding to eliminate the lags that were not relevant. Our method is provably consistent for a wide range of conditions \citep{Ost&Takahashi}.

The \pkg{hdMTD} package also implements a computationally efficient perfect (exact) sampling algorithm for MTD models. The main advantage of the perfect sampling algorithm is that it does not require choosing hyperparameters like burn-in duration, and it is not necessary to verify whether the obtained sample is stationary, as it is guaranteed to generate stationary samples. The exact sampling algorithm is especially relevant when the order of the MTD is large, as choosing hyperparameters or verifying the stationarity of the sample with long-range dependence can be hard. Perfect sampling algorithm for Markov chains of order $d$ is a theoretically well-studied topic, but a practical implementation seems to be lacking \citep{Comets2002Proceses}. To our knowledge, \pkg{hdMTD} is the first implementation of perfect simulation specific to MTD models. We expect that the toolset included in \pkg{hdMTD} will help with the analysis of stochastic phenomena with long-range dependencies that were previously impervious to rigorous statistical analysis.

\section*{Acknowledgments}
M.G. was supported by a PhD scholarship from CAPES (Coordenação de Aperfeiçoamento de Pessoal de Nível Superior – Brazil). 
She also received support from FAPERJ through the “Mestrado Nota 10” (MSC-10) fellowship during her MSc studies (grant E-26/200.454/2021), when the \pkg{hdMTD} package was first developed.
G.I.~was supported by FAPERJ (grant E-26/210.516/2024).
G.O. was supported by the Serrapilheira Institute (grant Serra – 2211-42049), FAPERJ (grant E-26/204.532/2024) and CNPq (grant 303166/2022-3).
D.Y.T. was partially supported by Serrapilheira grant R-2401-47364, CNPq Grant 421955/2023-6, and UFRN Grant Apoio a Eventos Interligados.

\bibliographystyle{plainnat}
\bibliography{refs}

\newpage

\end{document}